\documentclass[alpha-refs]{wiley-article}

\usepackage{siunitx}
\usepackage{booktabs} 
\usepackage{multirow}
\usepackage{xcolor}

\newenvironment{algocolor}{%
   \setlength{\parindent}{0pt}
   \itshape
   \color{black}
}{}

\papertype{Research Article}
\paperfield{ }

\title{GeoAnalystBench: A GeoAI benchmark for assessing large language models for spatial analysis workflow and code generation}

\author[1]{Qianheng Zhang}
\author[1]{Song Gao *}
\author[1]{Chen Wei}
\author[1]{Yibo Zhao}
\author[1]{Ying Nie}
\author[2]{Ziru Chen}
\author[2]{Shijie Chen}
\author[2]{Yu Su} 
\author[2]{Huan Sun}

\affil[1]{Geospatial Data Science Lab, \\Department of Geography, University of Wisconsin-Madison, Madison, WI, 53706, United States}
\affil[2]{Department of Computer Science and Engineering, The Ohio State University, Columbus, OH, 43210, United States}

\corraddress{Song Gao, PhD, Department of Geography, University of Wisconsin-Madison, Madison, WI, 53706, United States}
\corremail{song.gao@wisc.edu}

\fundinginfo{NSF, Award Number: 2112606}

\runningauthor{Zhang et al.}

\begin{document}

\begin{frontmatter}
\maketitle

\begin{abstract}
Recent advances in large language models (LLMs) have fueled growing interest in automating geospatial analysis and GIS workflows, yet their actual capabilities remain uncertain. In this work, we call for rigorous evaluation of LLMs on well-defined geoprocessing tasks before making claims about full GIS automation. To this end, we present GeoAnalystBench, a benchmark of 50 Python-based tasks derived from real-world geospatial problems and carefully validated by GIS experts. Each task is paired with a minimum deliverable product, and evaluation covers workflow validity, structural alignment, semantic similarity, and code quality (CodeBLEU). Using this benchmark, we assess both proprietary and open source models. Results reveal a clear gap: proprietary models such as ChatGPT-4o-mini achieve high validity (95\%) and stronger code alignment (CodeBLEU 0.39), while smaller open source models like DeepSeek-R1-7B often generate incomplete or inconsistent workflows (48.5\% validity, 0.272 CodeBLEU). Tasks requiring deeper spatial reasoning, such as spatial relationship detection or optimal site selection, remain the most challenging across all models. These findings demonstrate both the promise and limitations of current LLMs in GIS automation and provide a reproducible framework to advance GeoAI research with human-in-the-loop support.

\end{abstract}
\end{frontmatter}

\section{Introduction}

Recent advances in geospatial artificial intelligence (GeoAI) have been driven by generative AI and foundation models such as large language models (LLMs), vision-foundation models, and multimodal foundation models~\citep{jakubik2023foundation,wu2023mixed,huang2024zero,mai2024opportunities,wang2024gpt}. LLMs function as language assistants, copilots, and agents, enabling various types of users to interact with complex systems through natural language. Numerous studies have examined the abilities of LLMs in planning, reasoning, and solving complex tasks across multiple science domains as independent agents~\citep{chen2024scienceagentbench,yue2024mmmu}. Within the domain of Geographic Information Systems (GIS), researchers have investigated the ability of LLMs to understand inputs of spatial relationships between geographic objects and generate various visual representations in formats such as maps, graphs, and Well-Known Text (WKT) outputs~\citep{zhang2024mapgpt,ji-2023}. These studies indicate a continuous improvement in LLMs' capabilities, demonstrating that advancing automation of complicated GIS workflows to collect, analyze, and visualize GIS data is promising~\citep{li2023autonomous,zhang2024geogpt, wei2024geotool,jiang2024chatgpt}. In particular, LLMs have been shown to be effective assistants for individuals without prior knowledge of GIS or spatial contexts, lowering barriers to entry for spatial analysis by transforming traditional GIS workflows~\citep{zhang2024automating}. Although early studies are promising, there has not yet been a systematic test of whether LLMs can reproduce human expert-designed GIS workflows. Existing studies mostly focus on isolated tasks (e.g., generating a buffer or drawing a map), but few have examined whether LLMs can sequence multi-step geoprocessing operations, such as spatial joins, overlays, or raster calculations, in a coherent spatial analysis workflow aligned with expert logic. This leads to a fundamental question: \textbf{How do LLMs conceptualize and execute GIS tasks compared to human experts?} This fundamental question is systematically divided into three specific research questions to be answered in Section \ref{sec:results}.

\begin{itemize}
    \item  RQ1: How do proprietary LLMs perform compared to open-source models in geospatial workflow and code generation?
    \item  RQ2: What roles do domain knowledge (DK) and dataset description (DD) play in improving workflow generation accuracy?
    \item  RQ3: How do LLMs handle different aspects of code generation, including syntax, logical structure, and spatial data manipulation?
\end{itemize}

Beyond addressing technical challenges, the integration of LLMs into GIS workflows aims to democratize access to spatial analysis tools by a broader audience. Historically, GIS has been a domain that requires specialized knowledge, often creating barriers for people and communities with limited experience. Over the past decades, GIS technologies and tools have evolved significantly, improving the use of spatial data to inform decision making across various fields such as urban planning, environmental conservation, and disaster management~\citep{cervone-2015, pham2011case,andrew2015spatial}. However, these advancements have also introduced complexities, as researchers today can be confused with various data formats, programming environments, and software platforms~\citep{toms2015arcpy}. Using LLMs, these barriers can be mitigated, allowing a broader user community to engage with spatial data and derive actionable geospatial insights. More importantly, it has always been the wish for autonomous GIS analysts (GeoMachina) to become real towards artificial general intelligence (AGI) that can pass the Turing test ~\citep{janowicz-2019}.  Realizing this vision requires rigorous evaluation: Can LLMs reliably replicate the logical sequencing, parameter selection, and domain-aware reasoning that human experts apply to GIS tasks?

A key challenge in applying LLMs to geospatial workflows lies in their ability to replicate structured, domain-informed reasoning. Can LLMs determine a logical sequence of spatial operations? When addressing geospatial data analysis tasks, GIS experts typically decompose geoprocessing problems into logically ordered subtasks under a central topic, executing them in a logical sequence often based on domain knowledge and reasoning. For spatial data scientists, the primary workflow of finding optimal future fire stations includes: loading current fire stations and neighborhoods' data, buffering the location of current fire stations based on travel time/distance, and clipping them out to highlight the neighborhoods without sufficient fire station coverages. This sequential reasoning of workflow is intuitive to humans because each step builds on the previous one, forming a coherent geoprocessing workflow. In contrast, it remains unclear whether LLMs naturally follow the same structured reasoning process when solving spatial analysis problems with GIS tools. If a geoprocessing task requires spatial joins, geometric operations, and attribute queries, how does an LLM decide the sequence of operations? Furthermore, do LLMs understand the implicit relationships between data inputs and processing steps, such as recognizing that a JSON file stores point data in a GeoDataFrame format? Structured reasoning with logical sequences can help reduce hallucination in tools and parameter assignments while improving LLMs' reasoning abilities~\citep{wei2022chain}.  To systematically evaluate this capability, we introduce the \textit{GeoAnalystBench}, a benchmark designed to assess how closely LLM-generated workflows align with GIS expert-designed processes in both structure and logic.

\begin{figure}[!ht]
    \centering
    \includegraphics[width=0.98\linewidth]{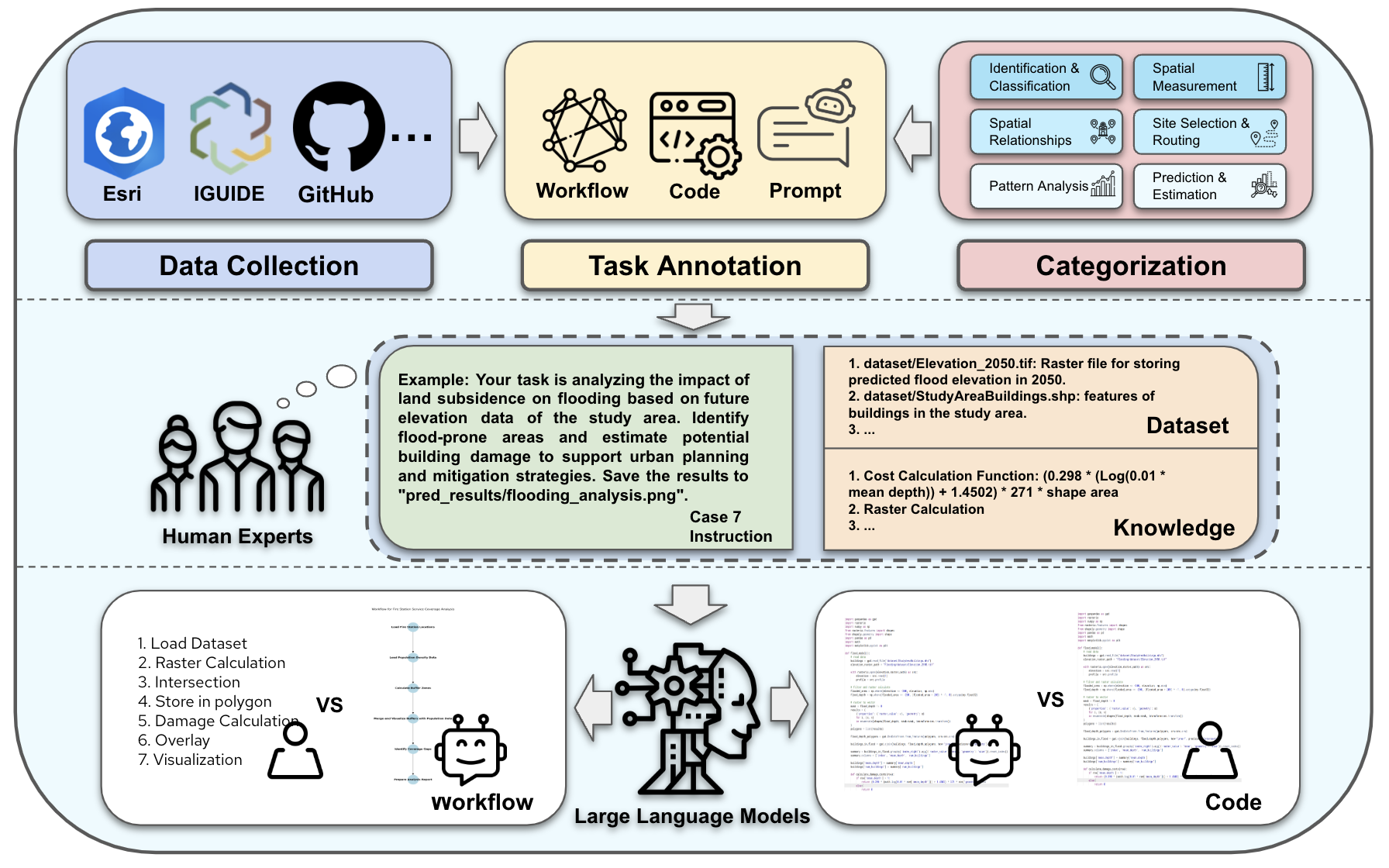}
    \caption{A conceptual framework for creating  geoanalyst benchmark datasets and evaluating the ability of LLMs to autonomously generate GIS workflows and code for spatial analysis tasks. }
    \label{fig:framework}
\end{figure}

To systematically address these challenges, this study presents a GeoAI benchmark dataset \textit{GeoAnalystBench} (\url{https://github.com/GeoDS/GeoAnalystBench}) to examine the abilities to solve spatial data science problems of LLM-powered AI agents compared to human experts. The benchmark dataset includes 50 major geoprocessing tasks derived from online GIS tutorials and academic publications, which focus on GIS experiments and exploratory spatial data analysis. Each major task can be divided into around 3 to 10 sub-level tasks. Then, concise human language prompts are passed to the LLMs for deriving the results. The ability of LLMs to generate geoprocessing workflows similar to those carefully designed by human experts was evaluated, with differences in workflow step length and structure being systematically measured.  GIS expert annotators then assess the quality of the workflow outputs, while tools such as CodeBLEU~\citep{ren2020codebleu} were used to evaluate the accuracy and effectiveness of LLM-generated code. By examining the alignment between human-designed and AI-generated workflows, this research provides a foundation to uncover the differences in logical structuring and explore whether these differences can be leveraged for improved geospatial analysis. Our findings will contribute to the design of more effective prompting strategies, future GIS automation tools, and hybrid GeoAI workflows that integrate foundation models and human thinking processes. Figure~\ref{fig:framework} illustrates the overall evaluation framework, including task prompting, code generation, execution, and expert comparison.

In summary, this study makes the following key contributions:

\begin{itemize}
    \item We develop the \textit{GeoAnalystBench}, a GeoAI benchmark dataset designed to systematically compare GIS workflows generated by LLMs with human-designed and annotated workflows.
    \item We analyze the logical structure of GIS tasks by LLMs, identifying key differences in performance across various GIS application domains.
    \item We evaluate LLM-generated codes using both \textit{human annotators} and automated metrics (e.g., \textbf{CodeBLEU}) to quantify the generated workflow accuracy and effectiveness.
    \item Our findings contribute to future development of improved prompting strategies, GIS automation tools, and hybrid human-AI workflows.
\end{itemize}

\section{Related Works}
The recent surge in LLM research has led to a variety of benchmarks and applications across domains, yet limited unified evaluation protocol exists for assessing LLM performance in spatially grounded reasoning and geoprocessing. Within GeoAI, researchers have focused on understanding how LLMs interpret spatial concepts, manage complex data formats, and produce meaningful spatial outputs.  This section reviews two complementary research directions: (1) the application or development of spatially-aware large language models tailored for GIS tasks and (2) GIS related benchmarks designed to systematically evaluate LLM performance in geospatial problem solving.

\subsection{Integration of Large Language Models with GIS}
A key strength of LLMs is their generalizability and adaptability, which allows them to be fine-tuned through prompt engineering and domain-specific training data. Recent works have taken such advantages to enhance GIS automation and spatial reasoning capabilities.

LLM-Geo~\citep{li2023autonomous} is a pioneering work on autonomous GIS and introduces a structured framework that decomposes spatial analysis tasks into discrete components: inputs, operations, intermediate data, and final outputs. These components are organized into a directed graph, enabling LLMs to reason through workflows step-by-step. This framework examines the chain-of-thought and planning capabilities of LLMs, emphasizing their ability to map complex geospatial queries into structured and executable operations. LLM-Geo's focus on workflow automation demonstrates the potential for LLMs to transition from general-purpose language models to domain-specific problem solvers with sophisticated planning and reasoning skills. 

Further advancing the integration of LLMs in GIS, GeoGPT~\citep{zhang2024geogpt} represents an early effort to utilize LLMs for solving spatial analysis tasks in GIS. This study introduces a framework for the generation, collection, and execution of GIS workflows via natural language prompts. This pioneering approach integrates LLMs with GIS tools to address end-to-end geospatial workflows and problem solving tasks, showcasing the potential of LLMs to serve as flexible assistants for GIS professionals. In a follow-up work,  a fine-tuned GeoTool-GPT was developed using annotated GIS instruction-solution evaluation data and demonstrated its effectiveness in operating GIS tools~\citep{wei2024geotool}. 
Building on this foundation, BB-GeoGPT~\citep{zhang2024bb} focuses on fine-tuning smaller LLMs to improve domain-specific performance. By using a curated dataset of GIS-related academic publications and Wikipedia articles, BB-GeoGPT enhances the performance in question-answering (QA) tasks across various GIS domains. This approach highlights how LLMs can be tailored to "understand" GIS as a subject, emphasizing the intersection of natural language and geospatial semantics. Unlike general-purpose models, BB-GeoGPT demonstrates how domain adaptation can bridge gaps in spatial reasoning and applications.  While these systems demonstrate the feasibility of integrating LLMs with geospatial tools, they typically focus on specific tasks or narrow capabilities, rather than providing a unified evaluation framework that systematically compares LLM performance across diverse GIS workflows and spatial reasoning steps.

 More recently, \cite{xu2025evaluating} introduced a multi‐task geospatial benchmark covering twelve spatial reasoning tasks, revealing substantial variability in LLM performance across domains. \cite{akinboyewa2025gis} developed GIS Copilot, an autonomous GIS agent integrated with QGIS that generates and executes spatial analysis workflows, achieving high success rates in tool selection and output correctness. \cite{krechetova-2025} proposed the GeoBenchX, a multistep geospatial task benchmark incorporating tool invocation and hallucination‐avoidance challenges. Complementing these, \cite{ning2025autonomous} designed an autonomous GIS agent framework for geospatial data retrieval, enabling LLMs to select data sources, generate code, and execute retrieval pipelines. \cite{li2025giscience} presented a research agenda towards autonomous GIS with LLMs and a conceptual framework that defines autonomous goals, autonomous levels, core functions, and operational scales of autonomous GIS. \cite{van2025opportunities} discussed opportunities and challenges of integrating GIS and LLMs. Collectively, these works demonstrate rapid advances in integrating LLMs with GIS and developing targeted benchmarks, while underscoring the need for unified, comprehensive evaluation protocols and social responsibilities. Additionally, \cite{janowicz2025geofm} provided a comprehensive review on how will LLMs and geo-foundation models reshape spatial data science and GeoAI.

\subsection{Benchmarks for the evaluation of LLMs in GIS}

While the above studies focus on the development of GIS-oriented LLMs, another line of research emphasizes systematic evaluation on benchmarks to uncover the pro and cons of current LLMs. 

For example, GeoGLUE~\citep{li-2023} introduces a geographic language understanding benchmark, which evaluates LLMs' abilities in geographic textual similarity searching, geographic element tagging, and geographic entity alignment, which are essential for understanding natural language and semantic reasoning within geographic contexts. This benchmark highlights the nuances of geographic expressions, including hierarchical naming and colloquial usage, offering diverse challenges for model evaluation on spatial relationships.

Focusing on multimodal reasoning, GeoQA~\citep{chen-2021} is a benchmark comprising 4,998 geometric problems, annotated with solving programs for transparency and evaluation. GeoQA demonstrates that LLMs achieve better results when textual and graphical inputs are combined. GeoQA establishes a robust benchmark sets a high standard for assessing numerical reasoning and multimodal data integration.

Focusing on code generation, \citet{gramacki2024evaluation} addresses the automation of geospatial analysis code generation in Python. This benchmark includes 20 unique tasks and 77 examples that are used to test various levels of complexity across GIS tools and data formats such as GeoPandas, ESRI Shapefiles, and GeoJSON. It finds that the performance of the LLMs' code generation is highly dependent on the difficulty of the task, the layout of the input, and the usage of the library. However, the model capabilities are constrained by the functions and libraries implemented within the benchmark, necessitating a broader tool integration. At the same time, GEE-OPs~\citep{hou2024gee} proposes the use of Google Earth Engine (GEE) platform on collecting 185,236 pieces of GEE code and using the Abstract Syntax Tree (which represent the structure of a code snippet) to extract the explicit operator relationships. The dataset noted with high-quality operator instructions improves LLMs' ability on understanding programming syntax as well as producing high-accuracy codes specific to GEE.

In sum, although several benchmarks have evaluated LLMs on geospatial tasks, most focus on textual understanding, question answering, or single-step operations. Recently, benchmarks such as GeoBenchX~\citep{krechetova-2025} and GEOBench-VLM~\citep{danish-2024} expand this landscape by tackling multi-step, multimodal, and tool-based reasoning challenges in geospatial tasks. ScienceAgentBench~\citep{chen2024scienceagentbench} offers 102 tasks across four scientific disciplines, including Geographical Information Science, with rigorous execution based evaluation on self contained Python programs, making it directly relevant as a GIS benchmark. Its task distribution explicitly includes geospatial analysis and map visualization subtasks, underscoring coverage of practical cartographic and analytical workflows. However, interpretability remains a challenge, especially when reasoning steps are opaque or hallucinated~\citep{singh-2024}. Our work differs by introducing a multi-step, execution-based benchmark that emphasizes workflow structure, tool chaining, and alignment with human expert logic. GeoAnalystBench thus addresses a unique evaluation gap at the intersection of code synthesis, spatial reasoning, and process automation.

\section{GeoAnalyst-Benchmark}
The complexity of real-world geospatial data science tasks often requires benchmark datasets to evaluate model performance on specific domains. To this end, this research developed 50 Python-based geoprocessing tasks derived from GIS platforms, software, online tutorials, and academic literature. Each task comprises 3 to 10 subtasks, because the simplest task still involves data loading, applying at least one spatial analysis tool, and saving the final outputs. The list of those tasks with their geoprocessing step lengths are included in the Table \ref{tab:tasks1-20}, Table \ref{tab:tasks21-40}, and Table \ref{tab:tasks41-50}.

The decision to focus on Python libraries stems from their suitability for automating GIS tasks. These tasks span various domains, including remote sensing, ecology, climate change, social science, and various application areas of spatial data science. Although the topics extend beyond traditional GIS, all tasks remain grounded in the practicalities of research experiences. Moreover, the tasks and their associated workflows and codes are designed to respect open source or copyrights generated by the annotators. By addressing a variety of scenarios, this research aims to evaluate how well the state-of-the-art LLMs can generate geoprocessing workflows and solutions for geospatial problems while highlighting the evolving role of these models in automation of spatial data science.

\subsection{Annotation Process
}

\begin{figure}[h]
    \centering
    \includegraphics[width=0.9\linewidth]{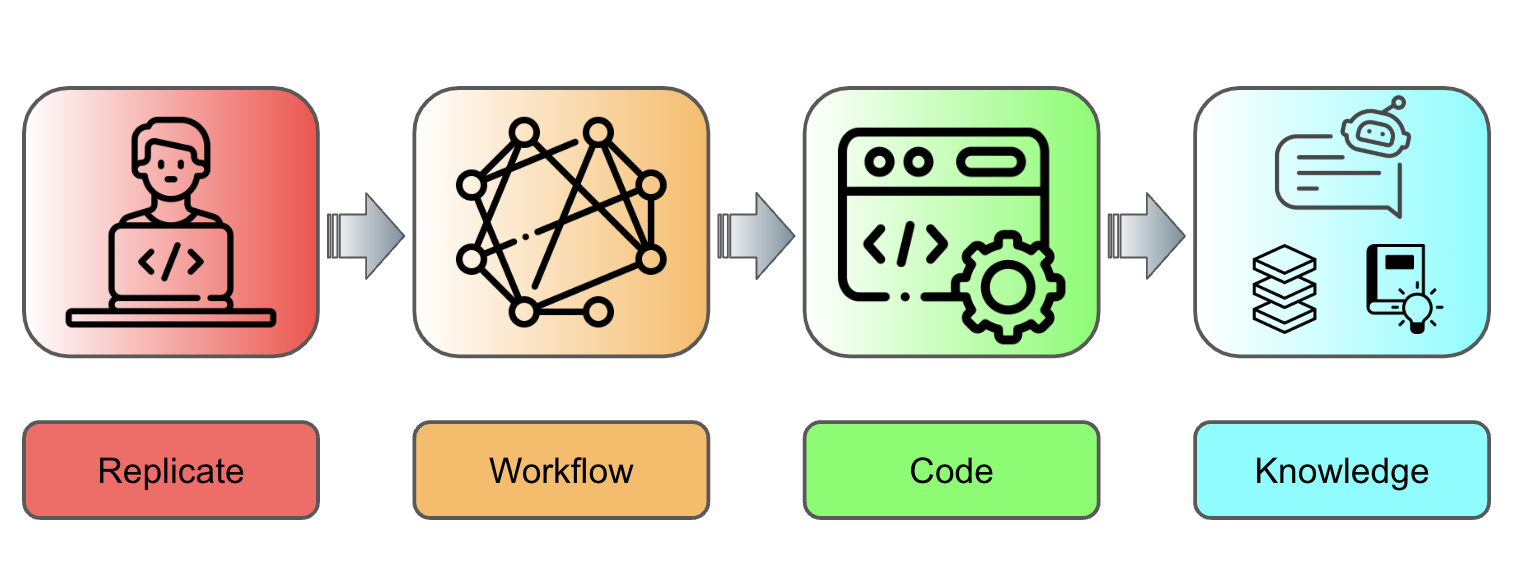}
    \caption{Annotators replicating the task, documenting workflow, programming the code, and consolidating into instructions, dataset descriptions, and domain knowledge.}
    \label{fig:annotate}
\end{figure}

To construct the benchmark tasks, we curated geospatial problem-solving examples from a combination of publicly available GIS tutorials (e.g., Esri Learn platform, university lab instructions) prior to year 2025. This approach aims to collect as many tasks as possible while providing high quality annotations. Each task was manually paraphrased, abstracted, and annotated by GIS experts to ensure distinct phrasing and structural independence from original sources. As shown in Figure~\ref{fig:annotate}, the annotators first carefully executed each task with GIS software such as ArcGIS or QGIS following the tutorials. Then, they systematically converted long tutorials into short-phrase formats of geoprocessing workflows. After that, these workflows will be translated into operational Python scripts. Finally, the annotators documented the essential domain knowledge and data description.  This multi-step curation process helps reduce circularity risk and enhances the benchmark’s validity for evaluating LLM reasoning. We also verified that the selected sources exhibit clear authorship and human editing history. A subset of tasks is derived from ESRI’s ModelBuilder examples, which produce deterministic outputs that serve as a basis for comparing LLM-generated code. Tasks include structured metadata and are designed to simulate authentic spatial analysis workflows without relying on memorized patterns.

To guide LLMs, each task was paired with a concise natural language instruction enriched with GIS domain-specific knowledge and detailed dataset descriptions. The instructions, on average around 100 words, outlined the desired methods or tools for spatial data processing and analysis. GIS domain knowledge clarifies specialized terms, algorithms, or classification methodologies, such as buffering, spatial statistics, or estimating property loss due to flood damage (case number 3, 8, 7 in table \ref{tab:tasks1-20}). This approach, similar to aspects of Retrieval-Augmented Generation (RAG)~\citep{lewis2020retrieval}, was simplified to improve its utility in evaluating LLMs' performance in different stages.

Dataset descriptions in Table \ref{tab:keyCols} provide critical context, detailing file formats (e.g., CSV, GeoJSON, GeoTIFF), column definitions, file sizes, and key attributes. For data files with multiple key feature columns, descriptions of the columns are specified because there are column names that are abbreviations or code names for specific features. This information addresses the complexity of the datasets and aimed to test whether equipping LLMs with detailed data descriptions could improve their ability to generate accurate workflows and solutions.

Three annotators read through the GIS online materials, replicated the processes outlined in the tutorials, and ensured that each task was suitable for automation using Python code. The instructions specify input files, intermediate tools, and final output formats, enabling evaluation of the LLMs' ability to connect inputs to outputs with the correct logical sequence and number of intermediate steps. This design facilitated the assessment of LLMs' capabilities in accurately understanding and replicating complex geoprocessing workflows.

The concise summary of knowledge in the \textit{GeoAnalystBench} is a modified version of \textit{ScienceAgentBench}~\citep{chen2024scienceagentbench}, which aims to test LLMs' abilities to solve data-driven scientific tasks in multiple science domains. This study refines the dataset description section to match the condition of spatial data layers. The approach in \textit{ScienceAgentBench} was trying to provide supplementary information to the LLM agents with first three rows of data, which doesn't fit to the large sized and abstract geospatial data. Because the geospatial data are relatively challenging to be informative in command-line style previews. Thus, this study only includes the data types, sizes, and key columns for the essential files. This approach aims to simulate the stage of exploratory data analysis stage for spatial analysis tasks. 

\begin{table}[!ht]
\centering
\caption{The Key columns and descriptions for GeoAnalystBench}
\label{tab:keyCols}
\begin{threeparttable}
\begin{tabular}{lp{10cm}}
\headrow
\thead{Key Columns} & \thead{Descriptions}\\
id & Unique identifier for each task\\
Open or Closed Source & Use open source or closed source library\\
Task & Brief description of the task \\
Instruction & Natural language instruction for completing the task \\
Domain Knowledge & Domain specific knowledge related to task \\
Dataset Description & Data name, format, descriptions, and key columns \\
Human Designed Workflow & Numbered list of human designed workflow \\
Task Length & The length of the human designed workflow \\
Code & Human designed code for the task and dataset\\
\hline  
\end{tabular}
\end{threeparttable}
\end{table}

\subsection{Spatial Analysis Categorization}
The \textit{GeoAnalystBench} dataset encompasses a wide range of spatial analysis topics while utilizing diverse Python toolkits. All tasks in this study are categorized as applied sciences of GIS and follow slightly modified ESRI's category definitions in the taxonomy of spatial analysis~\citep{esri-2013}:

\begin{enumerate}
  \item \textbf{Understanding where}: Tasks focused on identifying, classifying, and characterizing geographic locations and their attributes.
  \item \textbf{Measuring size, shape, and distribution}: Tasks that measure and evaluate the size, shape, orientation, and spatial distribution of geographic phenomena.
  \item \textbf{Determining how places are related}: Tasks that examine interactions, dependencies, and spatial correlations between different locations or geographic features.
  \item \textbf{Finding the best locations and paths}: Tasks related to spatial decision making, including site selection, route optimization, and spatial accessibility analysis.
  \item \textbf{Detecting and quantifying patterns}: Tasks focused on detecting spatial clusters, trends, anomalies, and spatial heterogeneity within datasets.
  \item \textbf{Spatial interpolation and predictive modeling}: Tasks that estimate unknown values at unsampled locations using spatial data, as well as forecasting spatiotemporal trends based on historical patterns.
\end{enumerate}

This categorization organizes spatial analysis/geoprocessing tasks based on their methodological and domain-specific approaches, ensuring both functional coherence and technical consistency. Given that all tasks in this benchmark are designed to simulate real-world scenarios with well-defined geospatial workflows within ten steps, it is crucial to adopt the applied science definitions to evaluate LLM performance from multiple perspectives.

\textbf{Understanding where} involves identifying the relative locations of geographic objects within a given space, facilitating spatial awareness and navigation. 
\textbf{Measuring size, shape, and distribution} extends beyond simple measurements to incorporate both quantitative properties and abstract spatial patterns. 
\textbf{Determining how places are related} focuses on identifying shared attributes and interconnections between multiple locations within the same space. 
\textbf{Detecting and quantifying patterns} involves analyzing recurring spatial structures, enabling the classification and grouping of related geographic entities. 
\textbf{Finding the best locations and paths} pertains to spatial optimization tasks, including site selection, route planning, and network analysis to determine optimal configurations based on predefined criteria. 
\textbf{Spatial interpolation and predictive modeling} involve estimating unknown values in spatial datasets or forecasting future trends based on spatial dependencies and historical data patterns.

\begin{figure}[!ht]
    \centering
    \includegraphics[width=0.9\textwidth]{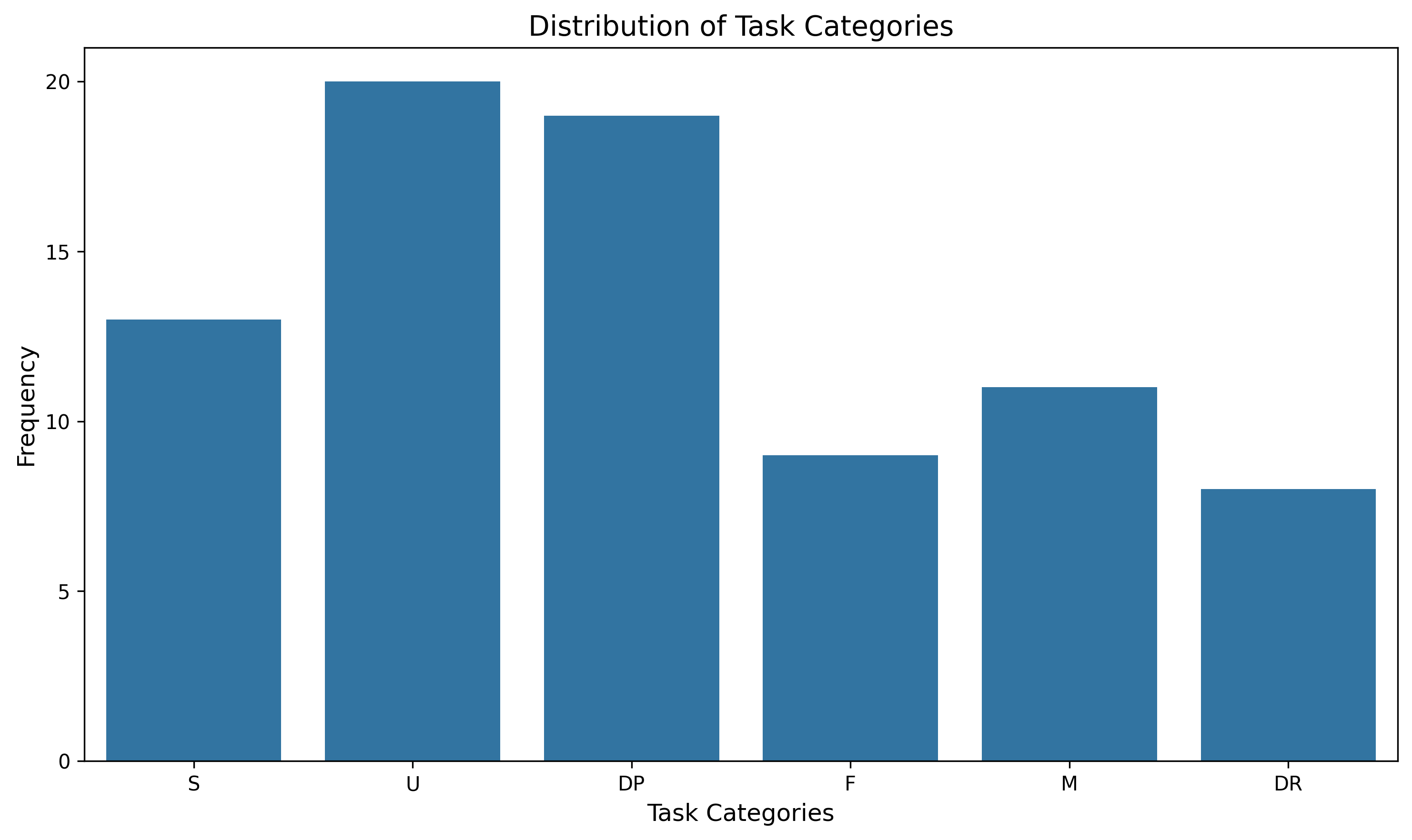} 
    \caption{The distribution of task categories included in the GeoAnalystBench: \textbf{U}: Understanding where; \textbf{M}: Measuring size, shape, and distribution; \textbf{DR}: Determining how places are related; \textbf{F}: Finding the best locations and paths; \textbf{DP}: Detecting and quantifying patterns; \textbf{S}: Spatial interpolation and predictive modeling.}
    \label{fig:taskcategory}
\end{figure}

This categorization simplifies the identification of tasks that share methodological similarities, such as spatial analysis, machine learning, or data visualization. Additionally, tasks may belong to at most three different categories to reflect their complexity and enhance diversity. By employing this structured approach, this study ensures a comprehensive coverage of spatial analysis/geoprocessing methodologies while maximizing the applicability of the benchmark in a wide range of real-world scenarios. At the same time, it's critical for unveiling the performance of LLMs on different aspects of spatial analysis. The topic distribution of the spatial analysis/geoprocessing tasks included in the \textit{GeoAnalystBench} is shown in Figure~\ref{fig:taskcategory}. 

\section{Evaluation Metrics}

Evaluating this benchmark dataset across various LLMs enables a comprehensive understanding of the key capabilities of LLMs as autonomous agents in solving spatial analysis/geoprocessing tasks: \textbf{workflow generation} and \textbf{code generation}. Since the task instructions are provided in natural language without explicitly specifying the GIS tools or programming languages, this benchmark dataset allows evaluation under different experimental settings. In our study, we specify \textit{Python} as the primary programming language for execution.

For each task, we generate four different prompts: \textit{task alone}, \textit{task with domain knowledge}, \textit{task with dataset descriptions}, and \textit{task with extra information}. The responses are generated using an online API from three proprietary LLMs (i.e., ChatGPT-4o mini, Claude 3.5 Sonnet, and Gemini 1.5) and three open-source ones (i.e., Llama 3.1, CodeLlama, and DeepSeek-R1). Three outputs per prompt are collected for evaluation.

The evaluation framework was implemented using both cloud-based APIs and local inference environments.  Proprietary LLMs such as ChatGPT 4o mini(2024-07-18), Claude 3.5 Sonnet(2024-6-20), and Gemini 1.5 Flash(2024-5-14) were accessed through their official APIs with a temperature setting of 0.7 to balance consistency and response diversity. Open-source models including Llama 3.1 8B(2024-7-23), DeepSeek R1 7B(2025-1-20), and CodeLlama 7B(2023-7-18) were run locally using the vLLM inference engine on a workstation with an NVIDIA RTX 2080 Super GPU and an Intel Core i7 processor. All models generated geoprocessing workflows and ArcPy code in response to structured prompts. These outputs were assessed using evaluation metrics such as task validity, mean absolute deviation in workflow length, and CodeBLEU scores. A separate system with an NVIDIA Tesla P40 GPU was used only for visualizing case study examples and validating reference code. This setup ensures a controlled and reproducible environment for performance comparison across model types.

In designing the benchmark, certain task instructions were intentionally left with ambiguous parameters (such as buffer size or clustering thresholds) to evaluate the LLMs’ ability to make reasonable assumptions and defaults in the absence of explicit guidance. This decision reflects realistic user behavior, where non-expert users may issue incomplete or underspecified queries. Allowing models to resolve such ambiguities provides insight into their domain intuition, default-setting behavior, and overall robustness. For example, if a prompt asks to "buffer schools in an urban area" without specifying the distance, the model’s choice reveals how well it understands standard practices in urban GIS applications. These domain-informed assumptions and decisions are then evaluated based on how closely they align with expert-designed workflows or plausible GIS conventions. While we do not explicitly label or isolate these tasks in the current study, this dimension of prompt ambiguity offers a promising direction for future research on LLM reasoning under uncertainty.


\subsection{Workflow Evaluation}

The \textit{workflow evaluation} assesses how well the generated geoprocessing workflows align with human-designed workflows, reflecting the ability of LLMs to structure geospatial problem-solving processes. Three key dimensions are considered:

\begin{itemize}
    \item \textbf{Validity} – The percentage of valid responses relative to the total number of LLM responses. A response is considered valid if it contains an extractable workflow list. Higher valid rates indicate better abilities to follow instructions and maintain coherence. In our implementation, validity is assessed through a combination of automated parsing and manual checking to ensure that the generated workflow constitutes a coherent and complete sequence of geoprocessing steps that satisfy the task instruction.
    
    \item \textbf{Complexity} – Measured by the length of the workflow, indicating how LLMs structure problem-solving steps. Most tasks are solvable within \textbf{6 steps}, making workflow length a critical factor in evaluating the efficiency of LLM design.  We use \textit{Mean Absolute Deviation (MAD)} to quantify how much the LLM-generated workflow length deviates from the human-designed reference. It is calculated as:
    
    \begin{algocolor}
    \[
    \text{MAD} = \frac{1}{N} \sum_{i=1}^{N} \left| L_{\text{LLM}}^i - L_{\text{human}}^i \right|
    \]
    \end{algocolor}
        
 where $L_{\text{LLM}}^i$ and $L_{\text{human}}^i$ represent the number of steps in the $i$-th workflow generated by the LLM and the expert, respectively. A smaller MAD suggests a closer match to expert logic, while a larger value may indicate that the LLM either overcomplicates or oversimplifies the task. In this study, MAD serves as an interpretable and task-agnostic measure of structural alignment between human and machine reasoning.
    
    \item \textbf{Text Similarity Score} – Assessed through \textit{text similarity scores} using pre-trained sentence transformers. The all-MiniLM-L6-v2 model, a fine-tuned version of MiniLM~\citep{Wang_Wei_Dong_Bao_Yang_Zhou_2020}, encodes the workflows in semantic embeddings and we use the embeddings to calculate the cosine similarity. The cosine similarity score results within 1 and -1, when a higher value indicates a higher similarity. This approach evaluates how closely machine-generated workflows align with human-created workflows in terms of semantic similarity. A high and positive similarity score suggests that the LLM's decisions successfully align with expert reasoning and step dependencies.
\end{itemize}

By integrating these three dimensions, the evaluation framework aims to assess (1) how valid the LLM-generated workflows are, (2) whether LLMs can replicate GIS experts' decision making in structuring spatial analysis/geoprocessing tools as solutions, and (3) maintaining workflow efficiency.
The following prompt template limits LLMs to output codes and tasks that are easy to visualize and process.

\begin{verbatim}
#Workflow Generation Prompt Templates
[Task Instructions]

[key Notes]
1.Use **automatic reasoning** and clearly explain each step.
2.Using **NetworkX** package for workflow graph visualization.
3.Using **Graphviz** `dot' for graph visualization layout.
4.Multiple subtasks can be proceeded simultaneously because
  all of their outputs will be inputs for the next subtask.
5.Limiting your output to code, no extra information.
6.Only codes for workflow generation, no implementation. 
\end{verbatim}

\subsection{Code Evaluation}

In addition, we also evaluate LLMs' capability in code generation for spatial analysis tasks. We employ the \textit{CodeBLEU}~\citep{ren2020codebleu}, an automatic code evaluation metric based on both bilingual evaluation understudy (BLEU) and abstract syntax trees (AST) in programming. The code evaluation includes the following aspects:

\begin{itemize}
    \item \textbf{N-gram match scores} - Comparing the occurrences of term pairs in the generated and reference code.
    \item \textbf{Weighted N-gram match scores} – Assigning different weights to N-gram matches according to importance.
    \item \textbf{Syntax AST match scores} – Assessing syntactic correctness in the programming code.
    \item \textbf{Semantic Data-flow match scores} – Evaluating whether the generated code manipulates the data in ways that are logically similar to human-written code.
\end{itemize}

These scores ensure that the evaluation captures both text similarity and syntactic correctness in Python programming languages. Considering the focus on data-driven spatial analysis workflow, we decide to use the final score with a weighted combination as shown in the following equation: 
\begin{equation}
\text{Score} = 0.2 \times \text{N-gram} + 0.2 \times \text{Weighted N-gram} + 0.3 \times \text{Syntax AST Match} + 0.3 \times \text{Semantic Data-flow Match}
\end{equation}

It is important to note that human-designed code, especially in the post-LLM era, may have been influenced by AI-generated solutions from online tutorials. To account for this, we compare \textit{AI-generated code} with \textit{human-designed code}, including code examples from both pre-LLM and post-LLM periods. At the same time, the machine-generated outputs generated from ArcGIS Pro are being compared for machine versus AI purposes.

\begin{figure}[!ht]
    \centering    \includegraphics[width=0.98\linewidth]{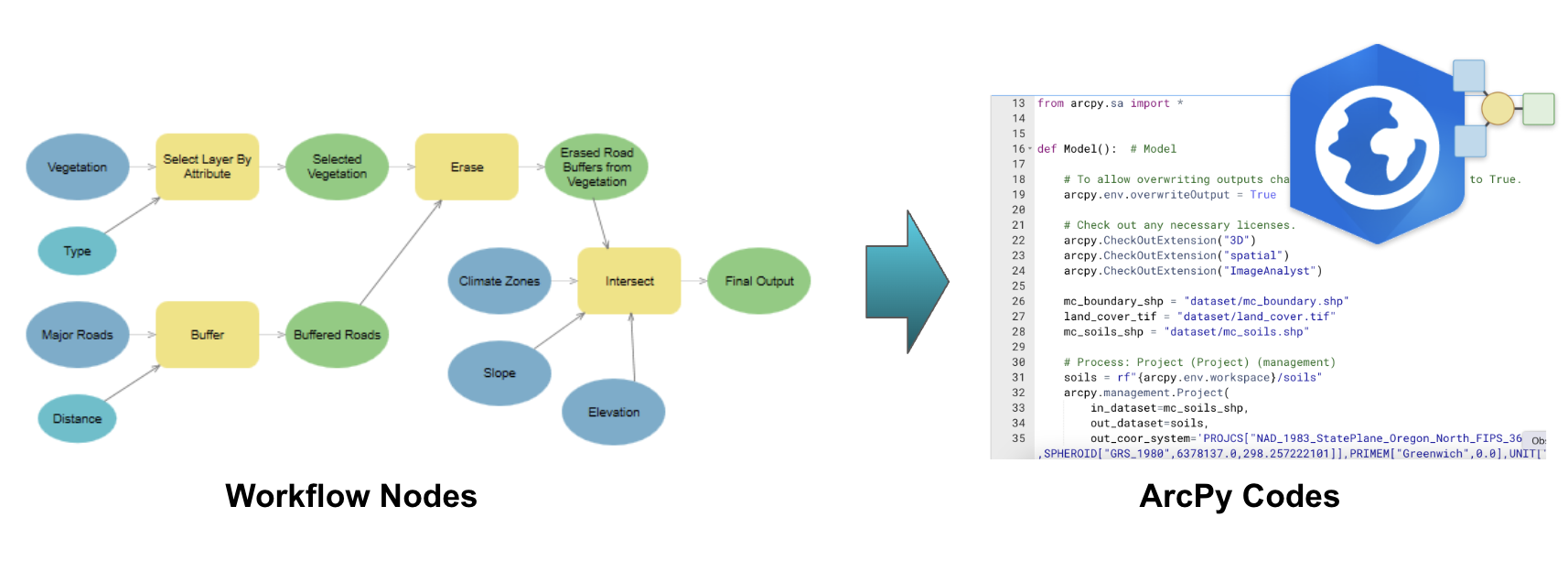}
    \caption{ModelBuilder allows users to add and connect data input, geoprocessing tools, and output as ArcPy Codes}
    \label{fig:modelBuilder}
\end{figure}

The tasks in the benchmark are collected not only from online resources, but also are crafted and verified by human experts. At the same time, ArcPy-oriented tasks employ a technique called \textit{ModelBuilder} that chains multiple geoprocessing tools in ArcGIS for solving spatial analysis tasks (in Figure \ref{fig:modelBuilder}). \textit{ModelBuilder} has a graphic user interface that allows users to visually connect geoprocessing tools/functions with specified data inputs and outputs. More importantly, the processing model can be output as python codes. These machine generated codes are more consistent than human-crafted codes, given that the \textit{ModelBuilder} always follows the same syntax on code format, which can serve as a solid format of ground-truth code during the evaluation.

The following prompt template limits LLMs to output codes that are easy to analyze.

\begin{verbatim}
#Code Generation Prompt Templates
[Task Instructions]

[key Notes]
1.Use **automatic reasoning** and clearly explain each subtask before performing it.
2.Using latest python packages for code generation.
3.Put all code under main function, no helper functions.
4.Limit your output to code, no extra information.
5.Use latest open source **Arcpy** packages only.
\end{verbatim}

\section{Results}
\label{sec:results}

This section presents how the 
\textit{GeoAnalystBench} is used to evaluate the performance of spatial analysis/geoprocessing workflow and code generation across various Large Language Models (LLMs) and LLMs-powered chatbots. These LLMs are categorized into proprietary and open-source groups based on their accessibility and development frameworks. The proprietary and closed-source models include ChatGPT-4o mini (20240718), Claude-3.5-Sonnet (20241022), and Gemini-1.5-Flash (20240924). These proprietary models are designed for general-purpose language understanding and reasoning. They are optimized for versatility, making them suitable for a wide range of tasks, including natural language processing and code generation. Among them, the Claude 3.5 Sonnet has demonstrated strong performance in zero-shot code generation tasks, as seen in prior benchmarks~\citep{anthropic2024claude35sonnet}. At the same time, open-source pretrained language models include DeepSeek-R1-7B and Llama-3.1-8B, which are among the most advanced publicly available LLMs. Llama 3.1-8B supports variety of natural language processing tasks, including reasoning and code generation~\citep{Grattafiori_Dubey_Jauhri_Pandey_Kadian_Al-Dahle_Letman_Mathur_Schelten_Vaughan_et_al._2024}. At the same time, DeepSeek-R1-7B is a knowledge distilled version of the larger DeepSeek-R1-671B model, which is created based on the Qwen-2.5 series models~\citep{yang2024qwen2} and fine-tuned by using 800k reasoning samples curated with the DeepSeek-R1-671B model.  The DeepSeek-R1-7B is small model but still with powerful reasoning abilities~\citep{DeepSeek-Ai_Guo_Yang_Zhang_Song_Zhang_Xu_Zhu_Ma_Wang_et_al._2025}. These models allow for modification and fine-tuning, enabling customization for geospatial applications. Additionally, we include the CodeLlama-7B, another open-source model specifically fine-tuned for code generation on Llama-2~\citep{Rozière_Gehring_Gloeckle_Sootla_Gat_Tan_Adi_Liu_Sauvestre_Remez_et_al._2023}. In addition, all models are evaluated with a temperature setting of 0.7 that keeps balance between creativity and consistency in their outputs.

\subsection{Workflow Generation}
\begin{table*}[!ht]
    \centering
    \caption{Performance evaluation of geospatial workflows generated by LLMs regarding their Valid Rates and Absolute Mean Difference for WK (Without Knowledge), DK (Domain Knowledge), DD (Dataset Description)}
    \label{tab:geoanalyst_performance}
    \renewcommand{\arraystretch}{1.1} 
    \setlength{\tabcolsep}{4pt} 
    \begin{tabular}{l|c|c|cc|cc|cc|cc}
        \toprule
        \multirow{2}{*}{\textbf{Models}} & \multicolumn{1}{c}{\textbf{VR}} & \multicolumn{1}{c}{\textbf{TS}}& \multicolumn{2}{c}{\textbf{WK}} & \multicolumn{2}{c}{\textbf{DK}} & \multicolumn{2}{c}{\textbf{DD}} & \multicolumn{2}{c}{\textbf{With Both}}\\
        & Rates & Scores & MAD & Std & MAD & Std & MAD & Std & MAD & Std \\
        \midrule
        \multicolumn{9}{l}{\textbf{Proprietary}} \\
        ChatGPT-4o-mini & 95.0 &  0.55 &\textbf{1.81} & \textbf{1.49} & \textbf{1.79} & \textbf{1.10} & \textbf{1.38}
        & \textbf{1.03} &  1.49 & 1.22 \\
        Claude-3.5-Sonnet & 93.5 &  0.54 & 2.09 & 1.50 & 1.90 & 1.53 & 1.82 & 1.46 &1.76 &1.42\\
        Gemini-1.5-Flash & \textbf{96.0} &  \textbf{0.56} & 2.30 & 1.76 & 1.89 & 1.58 & 1.70 & 1.47 & \textbf{1.33} & \textbf{0.97}\\
        \midrule
        \multicolumn{9}{l}{\textbf{Open Source}} \\
        DeepSeek-R1-7B & 48.5 & 0.23 & 3.26 & 2.05 & 3.96 & 2.24 & 4.41 & 2.06 & 4.61 & 2.20 \\
        Llama-3.1-8B & \textbf{95.3} &  \textbf{0.39} & \textbf{2.0} & \textbf{1.64} & \textbf{2.01} & \textbf{1.51} & \textbf{2.03} & \textbf{1.41} & \textbf{1.81} & \textbf{1.32} \\
        CodeLlama-7B & 32.7 & 0.08 & 4.88& 2.09 & 4.72 & 2.08 & 4.75 & 2.11 & 4.98 & 2.09 \\
        \bottomrule
    \end{tabular}
\end{table*}

To answer the research question RQ1 (see Introduction section), Table \ref{tab:geoanalyst_performance} demonstrates an evaluation of geospatial workflow generation in the \textit{GeoAnalystBench} across different LLMs, and it reveals the current gap between proprietary and open-source LLMs. We found that ChatGPT-4o-mini, Claude-3.5-Sonnet, and Gemini 1.5 Flash, consistently outperform the open-source LLMs in their valid rates (VR), mean absolute difference (MAD), and text similarity (TS) in generating geospatial workflows. In particular, Gemini-1.5-Flash achieves the highest valid rate (96\%) and TS score, while all three models have over 93\% valid rates and 0.54 TS score. At the same time, ChatGPT-4o-mini demonstrates the lowest MAD across multiple knowledge settings, indicating a balance between accuracy and consistency. These results may indicate that commercially pretrained LLMs are advantaged by large-scale and high-quality training data, state-of-the-art fine-tuning techniques, and parameter optimization and thus can be extremely effective for geospatial reasoning and workflow generation. Also, the lower standard deviations of these models show that they give more consistent outputs with less variation, an important quality for GIS use cases necessitating the reproducible outputs.

In contrast, small-sized open-source LLMs have a wide variation in performance, with Llama-3.1-8B being the only viable option compared to proprietary models. At a valid rate of 95.3\% and similar MAD values, Llama-3.1 indicates that open source LLMs are viable for geospatial workflow automation if additional refinements and domain-specific training strategies are implemented. However, models like DeepSeek-R1-7B and CodeLlama-7B show significant limitations on workflow generation, with valid rates as low as 32.7\% and high MAD values over 4.5, suggesting that their current architectures are less suitable for sophisticated geospatial reasoning tasks. 

A finer-grained analysis of DeepSeek-R1-7B indicates that its training via knowledge distillation is one of the main reasons for its poorer performance. Model distillation, which transfers knowledge from a larger teacher model to a smaller student model, is generally known to improve computational efficiency at the cost of decreased reasoning depth and generalization capacity~\citep{Hsieh_Li_Yeh_Nakhost_Fujii_Ratner_Krishna_Lee_Pfister_2023}. In contrast to fully trained end-to-end models with long multi-step reasoning trajectories, distilled models may approximate responses based on statistical correlations instead of strong, context-dependent logical computation. This is especially troublesome in geospatial analysis, where the tasks require hierarchical decision making, spatial-contextual sensitivity, and exact logical ordering. As a result, DeepSeek-R1-7B's poorer generalization capacity probably accounts for its lower valid rate (48.5\%) and larger MAD values for all experimental settings because it is not able to create consistent geospatial workflows that satisfy domain-specific logic. The full size of DeepSeek-R1 with 671 billion parameters is expected to have a better performance but would require resource-intensive computing hardware.

Furthermore, regarding the RQ2, the addition of domain knowledge (DK) and dataset descriptions (DD) greatly enhances accuracy across all proprietary LLMs, highlighting the value of structured contextual information for spatial analysis and GeoAI applications. The results imply that domain knowledge related to the task and dataset is beneficial to reduce the hallucination of models at all levels. Future work should prioritize further development of open-source models through targeted GIS corpora pretraining, optimization of inference strategies, and evaluation of computational efficiency to improve their suitability for practical spatial analysis scenarios.

\subsection{Code Generation}

To answer RQ3, we compared the CodeBLEU scores among varying LLMs to demonstrate an evident performance difference between open source and proprietary models in the generation of Python code for solving geospatial tasks in the \textit{GeoAnalystBench}. As shown in Table \ref{tab:geoanalyst_performance_tab2}, ChatGPT, Claude, and Gemini consistently outperform open source counterparts such as Llama, CodeLlama, and DeepSeek, with ChatGPT-4o-mini leading with the best average score (0.390), which shows to be highly stable under varying conditions. Claude-3.5-Sonnet (0.370) and Gemini-1.5-Flash (0.358) demonstrate a moderate performance, especially when given domain knowledge and dataset descriptions, which signals their strengths in using extra context well for structured code generation. DeepSeek-R1-7B (0.272) performs the worst among all the models. This suggests potential weaknesses of the knowledge distilled LLMs in context retention and multi-step reasoning, which are critical for geospatial workflow and code generation.

Further examination of the impact of domain knowledge and dataset descriptions reveals that most models improve when additional contextual information is provided, but the extent of improvement varies. Claude shows the highest adaptability, reaching a CodeBLEU score of 0.406 in the domain and dataset condition, while GPT remains consistently high performing across all categories. open source models also exhibit greater variability, with Llama and CodeLlama exhibiting moderate improvement, and DeepSeek exhibiting reduced performance. This variation indicates that proprietary models have been optimized based on more systematic programming data to better generalize. Given more geospatial context, they can perform better.

\begin{table*}[!ht]
    \centering
    \caption{Code evaluation results}
\label{tab:geoanalyst_performance_tab2}
    \renewcommand{\arraystretch}{1.1} 
    \setlength{\tabcolsep}{4pt} 
    \begin{tabular}{l|c|c|c|c|c}
        \toprule
        \multirow{1}{*}{\textbf{Models}} & \multicolumn{1}{c}{\textbf{CodeBLEU}} & \multicolumn{1}{c}{\textbf{N-gram}}& \multicolumn{1}{c}{\textbf{Weighted N-gram}} & \multicolumn{1}{c}{\textbf{AST Syntax}} & \multicolumn{1}{c}{\textbf{Data-flow}} \\
        \midrule
        \multicolumn{5}{l}{\textbf{Proprietary}} \\
        ChatGPT-4o-mini & \textbf{0.390} &  \textbf{0.077} &\textbf{0.107} & \textbf{0.563} & 0.282 \\
        Claude-3.5-Sonnet & 0.370 &  0.070 & 0.089 & 0.532 & 0.249 \\
        Gemini-1.5-Flash & 0.358 &  0.071 & 0.096 & 0.544 & \textbf{0.299} \\
        \midrule
        \multicolumn{5}{l}{\textbf{Open Source}} \\
        DeepSeek-R1-7B & 0.272 & 0.032 & 0.044 & 0.405 & 0.265 \\
        Llama-3.1-8B & \textbf{0.340} &  \textbf{0.040} & \textbf{0.054} & 0.471 & 0.252 \\
        CodeLlama-7B & 0.319 & 0.034 & 0.053 & \textbf{0.487} & \textbf{0.291} \\
        \bottomrule
    \end{tabular}
\end{table*}

In addition, the CodeBLEU component scores show that although LLMs produce syntactically correct code (average syntax score: 0.501), they have difficulties with lexical similarity (average n-gram: 0.0539, average weighted n-gram: 0.074) and logical correctness (average dataflow score: 0.274). The low n-gram scores imply that LLMs generate varied implementations instead of identical replicas of reference solutions, which can cause inconsistencies in standard geospatial workflows. However, the relatively high syntax score implies that most outputs are structurally correct, even if functionally imperfect. The moderate dataflow score further suggests that logical dependencies between variables and function calls are only partially preserved, highlighting the challenges in complex multi-step geospatial processing and reasoning. Improving geospatial fine-tuning and enforcing stricter logical structure could enhance both lexical accuracy and execution reliability in generated code for spatial analysis tasks.  Execution time can vary substantially across hardware and software environments, making it a less reliable metric for cross model comparison, especially in cases where many LLM generated workflows fail to run due to syntax or semantic errors. Instead, following insights from agent-based AI research~\citep{wei2022chain}, we emphasize reasoning quality and the logical sequencing of operations. Since execution efficiency has already been examined in ScienceAgentBench~\citep{chen2024scienceagentbench}, we focus here on spatial reasoning and GIS tool usage.

\subsection{Performance across Spatial Analysis Categories}

\begin{figure}[!ht]
    \centering
    \includegraphics[width=0.9\linewidth]{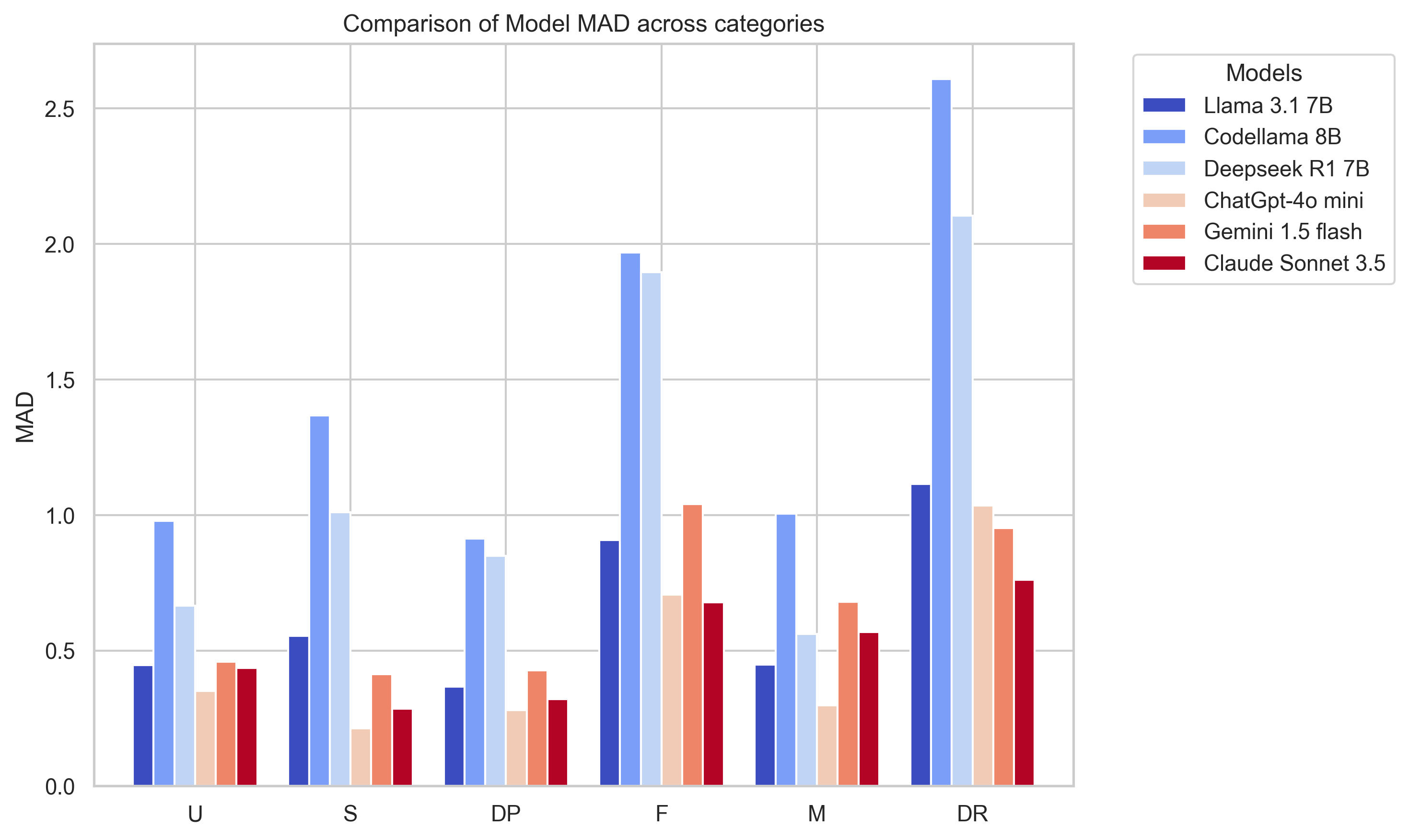}
    \caption{The mean absolute differences (MAD) across spatial analysis categories}
    \label{fig:mad}
\end{figure}

The performances of LLMs across different categories of spatial analysis tasks reveals a consistent gap between proprietary and open source models. As shown in Figure \ref{fig:mad}, ChatGPT, Gemini, and Claude demonstrate superior accuracy and lower MAD values across all the categories. Open source models, particularly DeepSeek-R1-7B and CodeLlama, struggle significantly with tasks requiring advanced spatial reasoning, such as determining spatial relationships (DR), finding optimal locations and paths (F), and predictive modeling (S). Llama-3.1-8B performs better than other open source alternatives but still lags behind proprietary models, particularly in spatial pattern detection and decision-making tasks. However, Llama-3.1-8B even outperforms Gemini and Claude in measuring size, shape, and distribution (M), indicating its competence in spatial measurement and geometric reasoning.

Both proprietary and open source LLMs perform worst in tasks of finding the best locations and paths (F) and determining how places are related (DR), indicating that these tasks involve more spatial reasoning and multi-step decision-making that the current LLMs are not able to perform well. The complexity of these tasks tends to include optimization, spatial network analysis, and hierarchical decision trees that may not be well captured in general-purpose LLM training data. LLMs do reasonably well at detecting and quantifying spatial patterns (DP) and measuring size, shape, and distribution (M) because these tasks depend on statistical pattern recognition and geometric reasoning. However, detecting spatial patterns (DP) and spatial interpolation (S) outcomes are highly disparate between different models, implying that certain proprietary LLMs may have been specifically optimized for pattern recognition tasks, but open source models like DeepSeek-R1-7B and CodeLlama are poor at predictive spatial modeling. The great disparity in spatial interpolation and predictive modeling indicates that these spatial analysis tasks need GIS domain-specific knowledge, structured training on good quality of datasets. The performance divergence supports the necessity of task-specific fine-tuning of LLMs and better spatially-aware embeddings or human-AI collaborative strategies to improve the performance in GIS-oriented spatial reasoning and location-based decision-making.

\subsection{Open versus Closed Source}

The above analyses reveal that ChatGPT, Claude, and Gemini consistently outperform other models across all evaluation criteria, making them strong candidates for GeoAI applications when integrating with LLMs. Open source models, in general, exhibit diverse performance across different spatial analysis tasks. In contrast, ArcPy-based outcomes demonstrate a better performance in domain knowledge and dataset evaluations, suggesting they are optimized for specific GIS workflows within controlled environments. Although Llama and DeepSeek lag behind overall performance, they still maintain respectable scores, indicating the potential for future improvement. The results suggest that ArcPy may be more suitable for specialized GIS tasks than open source packages. In summary, ChatGPT, Gemini, and Claude are the most balanced choices, presenting a trade-off between general AI and geospatial capabilities, which makes them promising tools that can significantly aid for GIS automation and sophisticated spatial analysis workflows.

\begin{figure}[!h]
    \centering
    \includegraphics[width=0.98\linewidth]{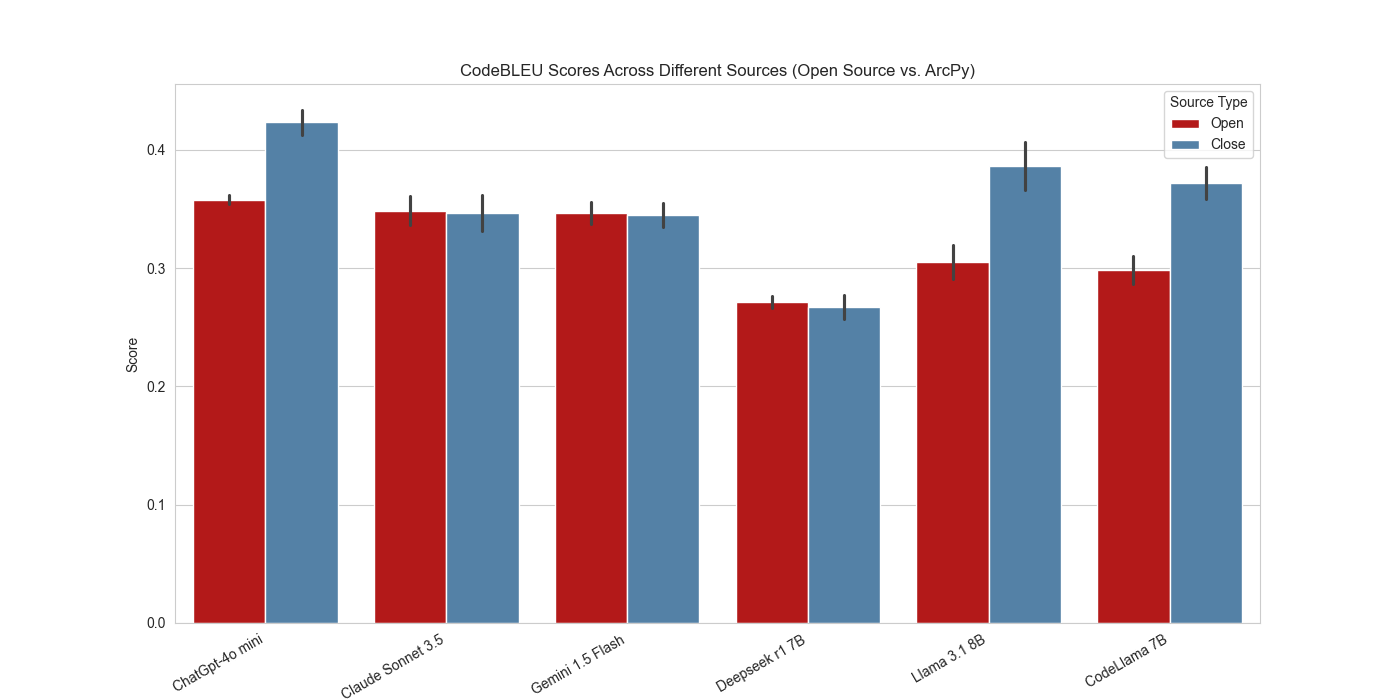}
    \caption{ArcPy versus open-sourced Packages on CodeBLEU Scores}
    \label{fig:ArcPyOpen}
\end{figure}

One possible reason behind the identified trend could be that the greater stability of ArcPy and its well-organized function references online might account for its better performance in the GIS domain-specific tasks. Since ArcPy is tightly integrated with the ArcGIS ecosystem, its standardized functions ensure consistent results across different processing workflows. Although ArcPy contains numerous tools, it has been well documented and are openly available online for user learning and pretraining of LLMs. This study found that ArcPy codes generated by ArcGIS ModelBuilders are similar to LLM generated code with relatively more stable structures and similar mindsets. In contrast, open source models rely on a variety of geospatial libraries such as GeoPandas, Shapely, and Rasterio, each with different syntax, functionalities, and update cycles. This diversity, while beneficial for flexibility and adaptability, may introduce inconsistencies in AI-generated workflows, potentially leading to minor variations in performance. Consequently, ArcPy's consistency probably makes it more efficient in advanced GIS operations, while open source packages, although having good overall performance, can be inconsistent in their performance due to the dispersed state of existing geospatial libraries.

\section{Case Studies}

In the following, we present two case studies included in the \textit{GeoAnalystBench} to demonstrate the capabilities of LLMs in geospatial workflow and code generation. 

\subsection{Case Study 1: Identification of Home Range and Spatial Clusters from Animal Movements}

\subsubsection{Background}
Understanding elk movement patterns is critical for wildlife conservation and management in the field of animal ecology. The task needs to identify elk home ranges in \textit{Southwestern Alberta, 2009} using GPS-tracking locations. In doing so, researchers are able to analyze their space use and movement clusters for elk populations. Understanding the home range of the elk population is essential for ensuring sustainability and stability of the wildlife.  This case study is based on the task 43 in Table \ref{tab:tasks41-50}, which is created based on ESRI's tutorial on modeling animal home range~\footnote{\url{https://learn.arcgis.com/en/projects/model-animal-home-range/}}.

\begin{figure}[!h]
    \centering
    \includegraphics[width=0.75\linewidth]{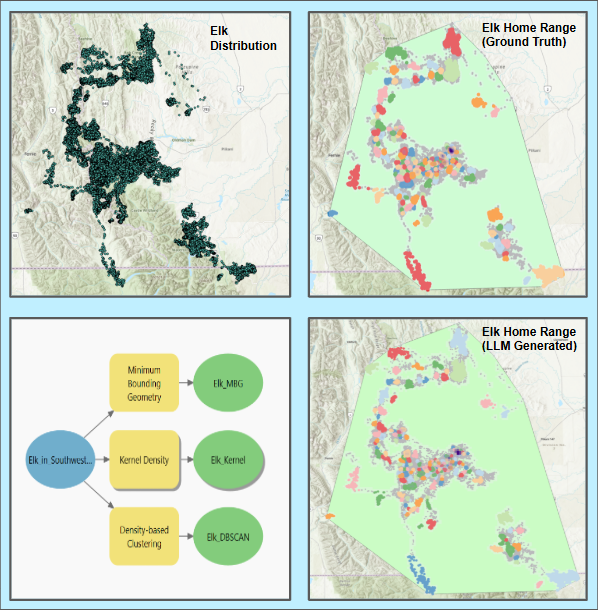}
    \caption{The geoprocessing workflow in ModelBuilder generates the ground truth result on the top right corner. The Gpt4o-mini generated code resulted in the right-bottom corner.}
    \label{fig:elkHomeRange}
\end{figure}

This case study takes GPS-tracked locations and applies three geoprocessing operations including: minimum bounding geometry, kernel density estimation, and density-based clustering to identify the home range, spatial concentration, and movement clusters of elk population groups, respectively. The final output is a map to overlay the three types of spatial analysis results. As a result, the image on the right in Figure \ref{fig:elkHomeRange} shows that the elk population is highly centric to the mountains and forest areas of the Piikani Reserve, while some separated elk groups extend their home range to places with human activities on the east side.



\subsubsection{Comparison of Human and LLM Approaches}
\vspace{0.5em}
To streamline the comparison, we present the expert-designed and LLM-generated workflows in a unified Table \ref{tab:elk_workflow_comparison}. Both approaches aim to analyze elk movement and home range, but differ in step granularity, tool usage, and output handling.
While both approaches utilize core geospatial methods—minimum bounding geometry, kernel density, and clustering—the LLM-generated workflow (see Appendix for detailed prompt design) includes additional preprocessing and visualization steps, emphasizing completeness and documentation over efficiency. In contrast, the human expert relies on ModelBuilder’s implicit visualization and streamlines the process with fewer, more targeted steps.

\begin{table}[!htbp]
\centering
\caption{Step-by-step Comparison of Human Expert vs. LLM Approach}
\label{tab:elk_workflow_comparison}
\begin{tabular}{|p{0.45\linewidth}|p{0.45\linewidth}|}
\hline
\textbf{Human Expert (ModelBuilder) Approach} & \textbf{LLM (ChatGPT-4o-mini) Approach} \\
\hline
Load dataset & Load GPS Data \\
\hline
Minimum Bounding Geometry (Convex Hull) & Preprocess Data (Clean \& Filter) \\
\hline
Kernel Density Estimation & Plot GPS Tracks \\
\hline
DBSCAN Clustering & Estimate Home Range (Convex Hull) \\
\hline
-- & Compute Kernel Density Estimation \\
\hline
-- & Apply DBSCAN Clustering \\
\hline
ModelBuilder auto-visualization & Analyze Home Range \& Habitat Preferences \\
\hline
-- & Store Outputs in Geodatabase \\
\hline
\end{tabular}
\end{table}

The workflow approaches in sections \textbf{6.1.2} and \textbf{6.1.3} are quite similar in methodology but with little difference in mindset. When the human approach is a concise GIS workflow, the LLM approach follows a more detailed structured pipeline. For example, it separates the loading and plotting process. while suggesting an analysis and storage process for better data organization. In this case, such approaches can be beneficial for GIS automation. This example also demonstrates the reason why we are considering both task lengths and text similarities: task length serves as a metric for how differently humans and LLMs reason and process in problem-solving workflows, while text similarity helps quantify the degree of alignment between their code outputs.

Evaluating the quality of LLM-generated workflows and codes remains a challenge. Although these models successfully outline the necessary steps for exploratory data analysis, their output often lacks optimization and high-quality refinement. As shown in Figure \ref{fig:elkHomeRange}, the result based on LLM-generated codes follows correctly the given instructions, successfully visualizing both the convex hull and the spatial clustering. However, the kernel density estimation (KDE) step employs a cell size twice of the optimized value in the LLM generated code, resulting in a low-resolution output. This discrepancy highlights the limitations of the LLMs in specific parameter selection, which can significantly affect the precision of spatial analyses.

The results also highlight a critical challenge: while user-defined parameter optimization is essential for refining spatial analysis outputs, the default settings used by LLMs often introduce randomness, which can be confusing for users unfamiliar with spatial analysis techniques. For instance, varying default clustering parameters (minimum points and search distance) in DBSCAN can lead to different interpretations of animal movement patterns. Non-expert GIS users may struggle to interpret these outputs without prior knowledge, making it difficult to refine or adjust the analysis effectively. This underscores the need for enhanced tuning mechanisms or built-in guidance within LLM-generated workflows to improve usability and interpretation. One potential solution is incorporating interactive parameter adjustments or AI-assisted explanations to help non-experts understand parameter choices and their implications in spatial analysis.

\subsection{Case Study 2: Spatial Hotspot Analysis of Car Accidents}
\subsubsection{Background}
The second case study is about spatial hotspot analysis of car accidents. The Brevard County in Florida has one of the deadliest interstate highways in the United States. This case study aims to identify the spatially distributed hot spots along the road network. The dataset includes road network, crash locations from 2010 to 2015, and a network spatial weighting matrix. Understanding the hot spots for car accidents is essential for the local transportation department to make policies and quick responses for future accidents.  This case study is based on the task 46 in table \ref{tab:tasks41-50} on ESRI's spatial analysis tutorial on determining the most dangerous roads for drivers~\footnote{\url{https://learn.arcgis.com/en/projects/determine-the-most-dangerous-roads-for-drivers/}}.

\begin{figure}[!h]
    \centering
    \includegraphics[width=0.9\linewidth]{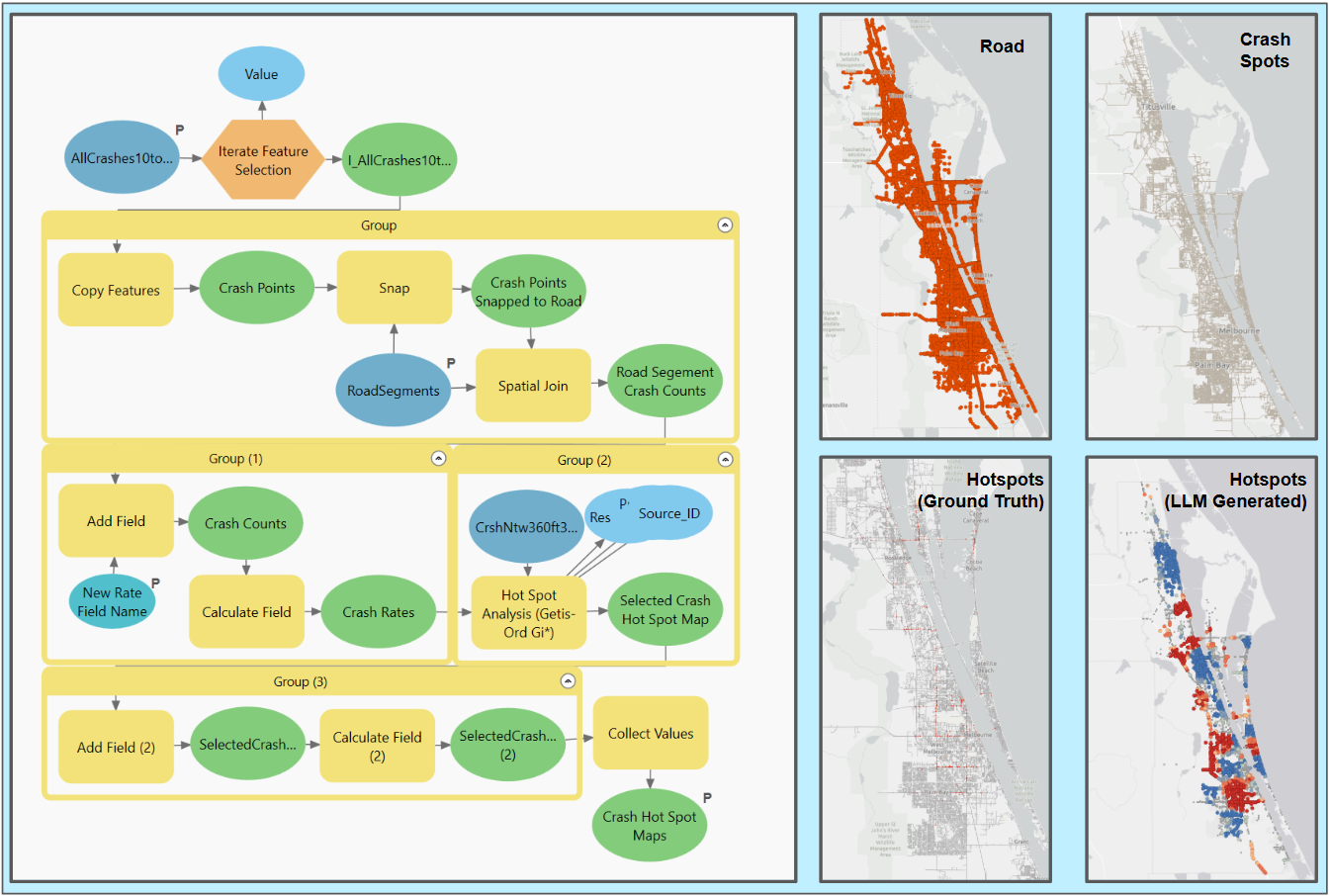}
    \caption{The left side persents the complicated workflow in the ArcGIS Pro Modelbuilder. Top right corner presents two data inputs. The first image on the bottom right corner is the human designed workflow's data. The second image is the spatial hotspot analysis result from the ChatGPT-4o-mini generated codes}
    \label{fig:TrafficAI}
\end{figure}

This task selects crash points to snap them onto roads, calculates the crash rates for each road, and creates a hot spot map using the spatial statistical analysis (Getis-Ord Gi*). In Figure \ref{fig:TrafficAI}, the hot spots show the car accidents tend to occur on major highways, bridges, and intersections, where traffic converges. In contrast, areas with intricate road networks and complex urban layouts might be considered cold spots in Brevard County.

\subsubsection{Comparison of Human and LLM Approaches}
\vspace{0.5em}
We summarize both the human-designed and LLM-generated workflows for crash hot spot detection in Table~\ref{tab:crash_workflow_comparison}. While the GIS expert relies on ArcGIS-specific tools that support reproducibility and precise execution, the LLM uses more generalized descriptions to represent each step.

\begin{table}[htbp]
\centering
\caption{Comparison of Human Expert vs. LLM Workflow for Crash Hot Spot Detection}
\label{tab:crash_workflow_comparison}
\begin{tabular}{|p{0.45\textwidth}|p{0.45\textwidth}|}
\hline
\textbf{Human Expert (ModelBuilder) Approach} & \textbf{LLM (ChatGPT-4o-mini) Approach} \\
\hline
Load dataset & Select peak-time crashes \\
\hline
Define crash hot spots based on time & Create copy of selected crashes \\
\hline
\textit{Copy Feature} to new layer & Snap crashes to road network \\
\hline
\textit{Snap} crash points to roads & Spatial join crashes with roads \\
\hline
\textit{Spatial Join} with road layer & Calculate crash rate per road segment \\
\hline
\textit{Add Fields} for crash count & Perform hot spot analysis \\
\hline
\textit{Calculate Fields} for crash rate & Generate crash hot spot map \\
\hline
\textit{Hot Spot Analysis (Getis-Ord Gi*)} & -- \\
\hline
\end{tabular}
\end{table}

\noindent
 Although both workflows follow similar core logic(filtering crashes, associating them with roads, and performing hot spot analysis), the human-designed version offers more operational granularity and tool specificity, making it immediately executable in ArcPy. The LLM-generated steps (see Appendix for detailed prompt design) abstract away from tool names, aiming for broader adaptability at the cost of precision.

\subsubsection{Result Comparision}

The second case study presents a more complex workflow compared to the first one, incorporating more GIS terminologies to describe the details tools/functions used. This complexity arises from the emphasis of the human editor on a transparent and tool-based workflow using ModelBuilder. In contrast, the LLM-generated approach in this case study is more abstract, focusing on generalizing the workflow rather than detailing individual tool applications. However, the human-designed workflow in the first case study tends to be more abstract and concise than its LLM-generated ones. This suggests that while human expertise adapts to different levels of details depending on the context, LLMs may follow a standardized approach.

Interestingly, as shown in Figure \ref{fig:TrafficAI}, the outcome from LLM-generated code in the second case study is significantly different from the GIS expert proposed approach while both workflows are quite similar. The impact of these differences is evident in the geovisual outputs. The human-designed workflow in Figure \ref{fig:TrafficAI} produces a road-based crash rate analysis, emphasizing the spatial patterns of car crashes along the road segments. This approach aligns with the transportation safety studies that focus on high-risk roadways. On the other hand, the LLM-generated Python script performs a point-based hotspot analysis, categorizing crash locations into statistically significant spatial hot and cold spots. This geovisualization provides a broader spatial context, highlighting crash-prone regions rather than specific road segments. The different results suggest that human experts prioritize structured spatial relationships between geographic features (e.g., crashes and roads), while LLMs generalize spatial patterns through statistical clustering. Both approaches follow the same instruction but diverge in focus due to differences in human understanding and LLMs' domain knowledge, particularly in data analysis strategies and the interpretation of spatial relationships for analytical purposes.

\section{Discussion}
 The comparative results from \textit{GeoAnalystBench} highlight several consistent patterns in how LLMs approach spatial analysis tasks. Proprietary models such as ChatGPT-4o-mini, Claude-3.5-Sonnet, and Gemini-1.5-Flash outperform open-source counterparts in valid rate, workflow alignment, and code quality. This performance advantage likely stems from their access to larger and more diverse training corpora, advanced instruction-tuning methods, and more extensive reinforcement learning from human feedback. These factors appear to improve not only syntax correctness but also the ability to incorporate domain knowledge and dataset descriptions into coherent geoprocessing pipelines.

 Differences in performance across spatial analysis categories provide insight into LLM generalization. Tasks in “Finding the best locations and paths” (F) and “Determining how places are related” (DR) consistently show higher difficulty, suggesting that multi-step optimization, spatial network analysis, and hierarchical decision-making require reasoning capabilities not yet well represented in general-purpose model training. In contrast, categories like “Measuring size, shape, and distribution” (M) or “Detecting and quantifying patterns” (DP) benefit from statistical pattern recognition skills that LLMs appear to generalize more effectively. This unevenness points to the need for GIS-specific pretraining that emphasizes relational and optimization-based reasoning.

 The results also reveal important differences within the open-source group. While smaller models such as DeepSeek-R1-7B and CodeLlama struggle to produce valid and efficient workflows, Llama-3.1-8B approaches proprietary performance in several scenarios. This indicates that with targeted fine-tuning on high-quality GIS corpora, open-source models could offer competitive alternatives for users requiring transparency and customizability.

From a practical standpoint, the evaluation metrics (valid rate, MAD, and CodeBLEU) used here map to critical qualities for real-world GIS automation. A high valid rate reflects reliability in producing runnable outputs; low MAD indicates workflow efficiency and avoidance of unnecessary complexity; higher CodeBLEU scores imply better preservation of logical and syntactic integrity, reducing the risk of execution errors. Together, these measures provide an interpretable framework for assessing robustness in operational settings, where reproducibility, efficiency, and correctness directly affect decision-making and project timelines.

 These findings suggest a pathway toward more usable LLM-powered GeoAI systems: leveraging proprietary model strengths where resources allow, while investing in domain-specific pre-training and fine-tuning to narrow the gap for open-source models. Integrating retrieval-augmented generation and human-in-the-loop validation may further improve generalization to complex, context-dependent spatial problems.

\section{Conclusion}
The results of this research showcase the strengths and weaknesses of the state-of-the-art Large Language Models (LLMs) in the context of spatial analysis workflows and GIS automation. By comparing LLM-generated workflows and code with GIS experts designed benchmarks, we offer important insights of these models in performing spatial analysis tasks. The most advanced LLMs have similar levels of performance for spatial analysis tasks, which demonstrates a powerful baseline for the current data-driven tasks in GeoAI. Importantly, spatial knowledge relate tasks (F and DR) are more difficult than regular data analysis tasks for the LLMs, indicating the limitations for current models' capabilities. At the same time, it's highly possible for fine-tuned LLMs to get better performance on spatial analysis tasks by adapting GIS software documentations through retrieval-augmented generation (RAG) techniques.

At the same time, the proposed \textit{GeoAnalystBench} is a valuable benchmark dataset for evaluating the workflow and code generation capabilities of LLMs on spatial analysis tasks. The benchmark builds on real-world datasets and domain-specific problems, each paired with a minimum deliverable product. While these tasks reflect complex, practical scenarios that can contribute to larger projects, their value also lies in probing whether foundation models approach problems similarly to human GIS analysts. This contrasts with recent work such as \textit{ScienceAgentBench}~\citep{chen2024scienceagentbench}, which shares several overlapping tasks but emphasizes the optimization of LLM-agent pipelines through domain expert feedback and detailed automation protocols. While both benchmarks explore LLM performance in structured scientific workflows, our focus is specifically on spatial analysis—a domain traditionally considered resistant to automation due to its complexity and reliance on geographic reasoning, spatial data formats, and context-aware interpretation. As such, \textit{GeoAnalystBench} offers a complementary perspective by evaluating how well LLMs can support spatial thinking and potentially democratize GIS practice for non-experts.

Despite the promising outcomes, there are several limitations that must be taken into account. First, the study focused on evaluating LLM-generated workflows for certain GIS tasks with predefined parameters, and thus the performance may vary when applied to new or more complicated real-world problems in different scientific domains. This process requires clearer definitions and high-quality annotations from domain experts. Second, tasks in the benchmark dataset varied in GIS tools being used, while limited by linear structures. In many cases, subtasks can be done in parallel, creating a different shape of workflow and generating multiple outputs. Furthermore, LLMs are also faced with data compatibility issues, such as incorrect parameter passing and incompatible data types, leading to execution failures. There is a great deficiency in the handling of spatial referencing and transformation, often leading to de-synchronized spatial data.  Some tasks were adapted from publicly available tutorials, so we can’t completely rule out the possibility that similar examples exist in LLM training data. As Figure \ref{fig:mad}, Table \ref{tab:geoanalyst_performance}, and Table  \ref{tab:geoanalyst_performance_tab2} suggested, the noticeable performance differences across models and spatial analysis categories suggest that the tasks remain sufficiently challenging and are not simply memorized. To overcome these deficiencies, it is necessary to enhance the LLMs with a good validation procedure and enrich their pretraining datasets with more diverse GIS-specific scenarios. 

Future research would focus on improving the spatial reasoning capabilities of LLMs in GIS by incorporating more specialized training datasets that include spatial analysis case studies. Currently, the \textit{GeoAnalystBench} is still limited by the coverage of real-world scenarios, leaving space for future improvement on the dataset side. At the same time, it is important to understand how the integration of the RAG technique will assist LLMs in automatically planning and organizing the geoprocessing tools. Interestingly, LLMs also have gaps between them. Furthermore, expanding the benchmark evaluation framework to include user studies would provide deeper insights into how GIS practitioners interact with LLM-generated workflows and what improvements are necessary to make them more effective in practical applications. The advancement in GeoAI research requires more community efforts on creating, maintaining, and assessing domain-specific benchmarks.

\section*{Acknowledgements}
We acknowledge the funding support from the National Science Foundation funded AI institute [Grant No. 2112606] for Intelligent Cyberinfrastructure with Computational Learning in the Environment (ICICLE). Any opinions, findings, and conclusions or recommendations expressed in this material are those of the author(s) and do not necessarily reflect the views of the funder(s).





\bibliographystyle{apalike}
\bibliography{references}


\clearpage

\renewcommand{\thefigure}{S-\arabic{figure}} 
\setcounter{figure}{0}  

\renewcommand{\thetable}{S-\arabic{table}} 
\setcounter{table}{0}  

\appendix
\section*{Appendix}
\section{Case Study 1: Identification of Home Range and Spatial Clusters from Animal Movements}
\label{appendix:elk_workflow}

\subsection{Workflow Prompt}

As a Geospatial data scientist, you will generate a workflow to a proposed task.

\begin{quote}
\textbf{[Task]}: Use animal GPS tracks to model home range to understand where they are and how they move over time.

\textbf{[Instruction]}: Your task is to analyze and visualize elk movements using the provided dataset. The goal is to estimate home ranges and assess habitat preferences using spatial analysis techniques, including Minimum Bounding Geometry (Convex Hull), Kernel Density Estimation, and Density-Based Clustering (DBSCAN). The analysis will generate spatial outputs stored in \texttt{dataset/elk\_home\_range.gdb} and \texttt{dataset/}.

\textbf{[Domain Knowledge]}: "Home range" can be defined as the area within which an animal normally lives and finds what it needs for survival. Minimum Bounding Geometry creates a feature class containing polygons that represent a specified minimum bounding geometry enclosing each input feature. "Convex hull" is the smallest convex polygon that can enclose a group of objects. "Kernel Density Mapping" calculates and visualizes feature density in a given area. "DBSCAN" (Density-Based Spatial Clustering of Applications with Noise) clusters the points based on a density criterion.

\textbf{[Dataset Description]}: \texttt{dataset/Elk\_in\_Southwestern\_Alberta\_2009.geojson}, containing GPS points of elk movements in Southwestern Alberta, 2009.

\textbf{Columns}: \texttt{'OBJECTID', 'timestamp', 'long', 'lat', 'comments', 'external\_t', 'dop', 'fix\_type\_r', 'satellite\_', 'height', 'crc\_status', 'outlier\_ma', 'sensor\_typ', 'individual', 'tag\_ident', 'ind\_ident', 'study\_name', 'date', 'time', 'timestamp\_Converted', 'summer\_indicator', 'geometry'}

\textbf{[Key Notes]}:
\begin{enumerate}
    \item Use automatic reasoning and clearly explain each step (Chain of Thoughts approach).
    \item Use \texttt{NetworkX} package for workflow visualization.
    \item Use \texttt{dot} layout for graph visualization.
    \item Subtasks may proceed in parallel; all outputs feed the next subtask.
    \item Limit output to code only.
    \item Only include workflow code, no implementation.
\end{enumerate}

\textbf{[Expected Sample Output Begin]}
\begin{verbatim}
tasks = [Task1, Task2, Task3]

G = nx.DiGraph()

for i in range(len(tasks) - 1):
    G.add_edge(tasks[i], tasks[i + 1])

pos = nx.drawing.nx_pydot.graphviz_layout(G, prog="dot")

plt.figure(figsize=(15, 8))
nx.draw(G, pos, with_labels=True, node_size=3000,
        node_color='lightblue', font_size=10,
        font_weight='bold', arrowsize=20)

plt.title("Workflow for Analyzing Urban Heat Using Kriging Interpolation", fontsize=14)
plt.show()
\end{verbatim}
\textbf{[Expected Sample Output End]}
\end{quote}

\subsection{Code Generation Prompt}

As a Geospatial data scientist, generate a Python file to solve the proposed task.

\begin{quote}
\textbf{[Task]}: Use animal GPS tracks to model home range to understand where they are and how they move over time.

\textbf{[Instruction]}: Analyze and visualize elk movements using the dataset. Estimate home ranges and assess habitat preferences using Minimum Bounding Geometry (Convex Hull), Kernel Density Estimation, and DBSCAN. Save outputs to \texttt{dataset/elk\_home\_range.gdb} and \texttt{dataset/}.

\textbf{[Domain Knowledge]}: Explained in workflow section above.

\textbf{[Dataset Description]}: \texttt{dataset/Elk\_in\_Southwestern\_Alberta\_2009.geojson}

\textbf{Columns}: Same as listed above.

\textbf{[Key Notes]}:
\begin{enumerate}
    \item Use automatic reasoning; explain each subtask before performing (ReAct approach).
    \item Use latest Python packages.
    \item Put all code under \texttt{main()}.
    \item Do not include helper functions.
    \item Limit output to code only.
    \item Use latest \texttt{arcpy} functions only.
\end{enumerate}
\end{quote}

\section{Case Study 2: Spatial Hotspot Analysis of Car Accidents}
\label{appendix:crash_hotspot_workflow}

\subsection{Workflow Prompt}

As a Geospatial data scientist, you will generate a workflow to a proposed task.

\begin{quote}
\textbf{[Task]}: Identify hot spots for peak crashes

\textbf{[Instruction]}: Your task is identifying hot spots for peak crashes in Brevard County, Florida, 2010–2015. The first step is to select all the crashes based on the peak time zone. Create a copy of the selected crash data. Then, snap the crash points to the road network and perform a spatial join with the roads. Calculate the crash rate from the joined data and use hot spot analysis to generate a crash hot spot map.

\textbf{[Domain Knowledge]}: Traffic between 3:00 pm and 5:00 pm on weekdays is considered peak time. For the snapping process, a buffer of 0.25 miles around roads is recommended. Hot spot analysis identifies spatial clusters with high crash rates, so accurate road-network-based distance measurements are crucial.

\textbf{[Dataset Description]}:
\begin{itemize}
    \item \texttt{dataset/crashes.shp}: Locations of crashes in Brevard County (2010–2015)
    \item \texttt{dataset/roads.shp}: Road network of Brevard County
    \item \texttt{dataset/nwswm360ft.swm}: Spatial weights matrix created from road polylines
\end{itemize}

\textbf{[Key Notes]}:
\begin{enumerate}
    \item Use automatic reasoning and clearly explain each step (Chain of Thoughts approach).
    \item Use \texttt{NetworkX} for visualizing the workflow.
    \item Use \texttt{dot} layout for graph visualization.
    \item Allow multiple subtasks to proceed in parallel if their outputs feed into the next step.
    \item Limit the output to code only, with no additional explanation.
    \item Only include workflow logic (no implementation).
\end{enumerate}

\textbf{[Expected Sample Output Begin]}
\begin{verbatim}
tasks = [Task1, Task2, Task3]

G = nx.DiGraph()

for i in range(len(tasks) - 1):
    G.add_edge(tasks[i], tasks[i + 1])

pos = nx.drawing.nx_pydot.graphviz_layout(G, prog="dot")

plt.figure(figsize=(15, 8))
nx.draw(G, pos, with_labels=True, node_size=3000,
        node_color='lightblue', font_size=10,
        font_weight='bold', arrowsize=20)

plt.title("Workflow for Peak Crash Hot Spot Detection", fontsize=14)
plt.show()
\end{verbatim}
\textbf{[Expected Sample Output End]}
\end{quote}

\subsection{Code Generation Prompt}

As a Geospatial data scientist, generate a Python file to solve the proposed task.

\begin{quote}
\textbf{[Task]}: Identify hot spots for peak crashes

\textbf{[Instruction]}: Your task is identifying hot spots for peak crashes in Brevard County, Florida, 2010–2015. First, filter crashes during the peak time zone. Create a copy of selected crashes. Snap points to roads, perform a spatial join, calculate crash rate, and apply hot spot analysis.

\textbf{[Domain Knowledge]}: Peak time is 3:00–5:00 pm on weekdays. Use a 0.25-mile buffer for snapping. Accurate road network distances are essential for meaningful hot spot analysis.

\textbf{[Dataset Description]}:
\begin{itemize}
    \item \texttt{crashes.shp}
    \item \texttt{roads.shp}
    \item \texttt{nwswm360ft.swm}
\end{itemize}

\textbf{[Key Notes]}:
\begin{enumerate}
    \item Use automatic reasoning (ReAct approach).
    \item Use the latest Python packages.
    \item Write all code inside a \texttt{main()} function.
    \item Do not use helper functions.
    \item Limit output to code only, with no explanatory text.
    \item Use the latest \texttt{arcpy} functions exclusively.
\end{enumerate}
\end{quote}

\newpage
\section{Supplementary Tables}
\begin{table}[!ht]
\caption{Geospatial Tasks (1–20). The table presents various GIS-related tasks along with their geoprocessing step lengths, access dates, and last update dates.}
\label{tab:tasks1-20}
\begin{threeparttable}
\begin{tabular}{l p{7.8 cm} r c c}
\headrow
\thead{ID} & \thead{Task Name} & \thead{Length} & \thead{Access Date} & \thead{Last Update} \\
1  & Find heat islands and at-risk populations in Madison, Wisconsin & 7 & 2024-7 & 2025-5 \\
2  & Find future bus stop locations in Hamilton & 6 & 2024-7 & 2025-1 \\
3  & Assess burn scars and wildfire impact in Montana using satellite imagery & 6 & 2024-7 & 2024-12 \\
4  & Identify groundwater vulnerable areas that need protection & 10 & 2024-7 & 2024-9 \\
5  & Visualize data on children with elevated blood lead levels while protecting privacy & 5 & 2024-7 & 2024-10 \\
6  & Use animal GPS tracks to model home range and movement over time & 6 & 2024-7 & 2025-1 \\
7  & Analyze the impacts of land subsidence on flooding & 7 & 2024-9 & 2025-7 \\
8  & Find gaps in Toronto fire station service coverage & 4 & 2024-6 & 2025-5 \\
9  & Find the deforestation rate for Rondônia & 6 & 2024-8 & 2025-1 \\
10 & Analyze the impact of proposed roads on the local environment & 5 & 2024-8 & 2025-1 \\
11 & Create charts in Python to explore coral and sponge distribution around Catalina Island & 7 & 2024-7 & 2023-10 \\
12 & Find optimal corridors to connect dwindling mountain lion populations & 3 & 2024-6 & 2025-1 \\
13 & Understand the relationship between ocean temperature and salinity at various depths in the South Atlantic Ocean & 6 & 2024-8 & 2025-7 \\
14 & Detect persistent periods of high temperature over the past 240 years & 4 & 2024-8 & 2025-7 \\
15 & Understand the geographical distribution of Total Electron Content (TEC) in the ionosphere & 3 & 2024-8 & 2025-7 \\
16 & Analyze climate change trends in North America using spatiotemporal data & 5 & 2024-8 & 2025-7 \\
17 & Analyze the geographical distribution of fatal car crashes in New York City during 2016 & 3 & 2024-8 & 2022-12 \\
18 & Analyze street tree species data in San Francisco & 4 & 2024-8 & 2022-12 \\
19 & Model spatial patterns of water quality & 4 & 2024-10 & 2025-6 \\
20 & Predict the likelihood of tin-tungsten deposits in Tasmania & 7 & 2024-8 & 2022-12 \\

\hline
\end{tabular}
\end{threeparttable}
\end{table}

\begin{table}[!ht]
\caption{Geospatial Tasks (21–40). The table presents various GIS-related tasks along with their geoprocessing step lengths, access dates, and last update dates.}
\label{tab:tasks21-40}
\begin{threeparttable}
\begin{tabular}{l p{7.8 cm} r c c}
\headrow
\thead{ID} & \thead{Task Name} & \thead{Length} & \thead{Access Date} & \thead{Last Update} \\
21 & Find optimal corridors to connect dwindling mountain lion populations(2) & 5 & 2024-6 & 2025-1 \\
22 & Find optimal corridors to connect dwindling mountain lion populations(3) & 3 & 2024-6 & 2025-1 \\
23 & Assess Open Space to Lower Flood Insurance Cost & 4 & 2024-7 & 2025-1 \\
24 & Provide a de-identified point-level dataset that includes all the variables of interest for each child, as well as their general location & 7 & 2024-10 & 2024-10 \\
25 & Create risk maps for transmission, susceptibility, and resource scarcity. Then create a map of risk profiles to help pinpoint targeted intervention areas & 10 & 2024-11 & 2025-2 \\
26 & Use drainage conditions and water depth to calculate groundwater vulnerable areas & 8 & 2024-11 & 2024-9 \\
27 & Identify undeveloped areas from groundwater risk zones & 6 & 2024-11 & 2024-9 \\
28 & Estimate the origin-destination (OD) flows between regions based on the socioeconomic attributes of regions and the mobility data & 7 & 2025-1 & 2025-1 \\
29 & Calculate travel time for a tsunami & 7 & 2024-7 & 2025-5 \\
30 & Designate bike routes for commuting professionals & 7 & 2024-7 & 2021-12 \\
31 & Detect aggregation scales of geographical flows & 7 & 2024-10 & 2019-6 \\
32 & Find optimal corridors to connect dwindling mountain lion population(4) & 8 & 2024-6 & 2025-1 \\
33 & Analyze the impacts of land subsidence on flooding & 5 & 2024-7 & 2025-7 \\
34 & Estimate the accessibility of roads to rural areas in Japan & 7 & 2024-12 & 2025-2 \\
35 & Calculate landslide potential for communities affected by wildfires & 10 & 2024-10 & 2024-6 \\
36 & Compute the change in vegetation before and after a hailstorm with the SAVI index & 7 & 2024-11 & 2025-7 \\
37 & Analyze of human sentiments of heat exposure using social media data & 6 & 2024-11 & 2024-4 \\
38 & Calculate travel time from one location to others in a neighborhood & 8 & 2024-11 & 2024-10 \\
39 & Train a Geographically Weighted Regression model to predict the population's Bachelor's degree rate in Georgia & 5 & 2024-11 & 2022-12 \\
40 & Calculate and visualize changes in malaria prevalence & 4 & 2024-11 & 2018-5 \\
\hline
\end{tabular}
\end{threeparttable}
\end{table}

\begin{table}[!ht]
\caption{Geospatial Tasks (41–50). The table presents various GIS-related tasks along with their geoprocessing step lengths, access dates, and last update dates.}
\label{tab:tasks41-50}
\begin{threeparttable}
\begin{tabular}{l p{7.8 cm} r c c}
\headrow
\thead{ID} & \thead{Task Name} & \thead{Length} & \thead{Access Date} & \thead{Last Update} \\

41 & Improve campsite data quality using a relationship class & 4 & 2024-11 & 2024-10 \\
42 & Investigate spatial patterns for Airbnb prices in Berlin & 8 & 2024-12 & 2024-8 \\
43 & Use animal GPS tracks to model home range to understand where they are and how they move over time (ArcPy Version) & 4 & 2024-7 & 2025-1 \\
44 & Find gap for Toronto fire station service coverage (ArcPy Version) & 5 & 2024-6 & 2025-5 \\
45 & Find optimal corridors to connect dwindling mountain lion populations (ArcPy Version) & 4 & 2024-6 & 2025-1 \\
46 & Identify spatial hot spots for peak crashes & 9 & 2024-12 & 2024-8 \\
47 & Calculate impervious surface area & 4 & 2024-12 & 2025-2 \\
48 & Determine how location impacts interest rates & 5 & 2024-12 & 2024-10 \\
49 & Mapping the Impact of Housing Shortage on Oil Workers & 4 & 2024-12 & 2023-8 \\
50 & Predict sea grass habitats & 6 & 2024-11 & 2025-5 \\
\hline
\end{tabular}
\end{threeparttable}
\end{table}
\end{document}